\documentclass[11pt]{article}

\usepackage[a4paper, margin=1in]{geometry}

\usepackage{amsthm,amsmath,amsfonts,amssymb}
\usepackage{lmodern}
\usepackage{microtype}

\usepackage[authoryear, round]{natbib}

\usepackage{xcolor}

\usepackage[colorlinks, citecolor=blue, urlcolor=blue, linkcolor=blue]{hyperref}
\usepackage{xurl}

\usepackage{graphicx}
\graphicspath{{figures/}{./figures/}}
\usepackage{float}
\usepackage{subcaption}
\usepackage{booktabs}
\usepackage{rotating}   

\usepackage[section]{placeins}  

\usepackage[linesnumbered, ruled, vlined]{algorithm2e}

\usepackage{tikz}
\usetikzlibrary{shapes,arrows,fit,calc,positioning}
\tikzset{box/.style={draw, diamond, thick, text centered, minimum height=0.5cm, minimum width=1cm}}
\tikzset{line/.style={draw, thick, -latex'}}

\usepackage{comment}
\usepackage{dsfont}
\usepackage{enumitem}
\usepackage{changepage}
\usepackage{bm}

\theoremstyle{plain}

\newtheorem{theorem}{Theorem}[section]

\newtheorem{assumption}{Assumption}
\newtheorem{proposition}{Proposition}

\theoremstyle{definition}
\newtheorem{definition}[theorem]{Definition}

\theoremstyle{remark}

\newcommand{\E}{\mathbb{E}}
\newcommand{\1}{\mathds{1}}
\newcommand{\p}{\mathbb{P}}

\newcommand{\ind}{\perp\!\!\!\perp}
\newcommand{\RCT}{\mathrm{RCT}}
\newcommand{\RWD}{\mathrm{RWD}}

\newcommand{\expit}{\operatorname{expit}}

\newcommand{\startsupplement}{%
  \clearpage
  \setcounter{section}{0}%
  \setcounter{equation}{0}%
  \setcounter{figure}{0}%
  \setcounter{table}{0}%
  \setcounter{proposition}{0}%
  \setcounter{corollary}{0}%
  \setcounter{algocf}{0}%
  \renewcommand{\thesection}{S.\arabic{section}}%
  \renewcommand{\thesubsection}{\thesection.\arabic{subsection}}%
  \renewcommand{\theequation}{S\arabic{equation}}%
  \renewcommand{\thefigure}{S\arabic{figure}}%
  \renewcommand{\thetable}{S\arabic{table}}%
  \renewcommand{\theproposition}{S\arabic{proposition}}%
  \renewcommand{\thecorollary}{S\arabic{corollary}}%
  \renewcommand{\thealgocf}{S\arabic{algocf}}%
}

\title{Bayesian fusion forests for heterogeneous treatment effects on survival from randomised and real-world data}

\author{
  Tijn Jacobs\thanks{Department of Mathematics, Vrije Universiteit Amsterdam. \texttt{t.jacobs@vu.nl}} \and
  St\'ephanie L.\ van der Pas\thanks{Department of Mathematics, Vrije Universiteit Amsterdam. \texttt{s.l.vander.pas@vu.nl}} \and
  Wessel N.\ van Wieringen\thanks{Department of Epidemiology and Data Science, Amsterdam University Medical Centre; and Department of Mathematics, Vrije Universiteit Amsterdam. \texttt{w.n.van.wieringen@vu.nl}}
}

\date{\today}

\begin{document}

\maketitle

\begin{abstract}
We develop the \emph{Bayesian fusion forest}, a nonparametric framework to estimate heterogeneous treatment effects on survival outcomes by combining a randomised controlled trial and real-world data.
The framework relaxes the unconfoundedness assumption on the real-world data by assuming instead that the treatment effect transports across the two sources.
Our method opens up right- and interval-censored outcomes to data fusion.
We model the survival time with an accelerated failure time decomposition into a shared baseline prognosis, a source-specific deviation, a treatment effect, and a confounding function.
The confounding function absorbs the confounding bias in the real-world data.
Each component receives a Bayesian tree ensemble prior.
The shared baseline prognosis borrows strength across sources, while the deviation captures between-source heterogeneity.
A hierarchical Dirichlet process mixture models the error distribution nonparametrically.
A simulation study shows efficiency gains over a trial-only analysis across varying levels of confounding and between-source heterogeneity.
We combine the ACTG~175 trial with the Multicenter AIDS Cohort Study to estimate the effect of combination antiretroviral therapy for HIV.
The fusion identifies a benefit for nearly every patient whereas the trial alone is inconclusive.

\medskip
\noindent\textbf{Keywords:} Bayesian additive regression trees; data fusion; heterogeneous treatment effects; real-world evidence; survival analysis; unmeasured confounding.
\end{abstract}



\section{Introduction}

Heterogeneous treatment effects are central to precision medicine.
Randomised controlled trials (RCTs) are the gold standard for causal effect estimation, but they are underpowered for subgroup effects and often have limited follow-up \citep{rothwell2005subgroup}.
Real-world data (RWD) from registries and cohorts are larger and often follow patients for longer \citep{sherman2016realworld}.
However, unmeasured confounding can bias treatment effect estimates from observational data.
We aim to combine the strengths of the two data sources: the trial identifies the treatment effect and the real-world data contribute power and longer follow-up.
Our motivating example is HIV antiretroviral therapy.
The treatment effect varies with baseline CD4 count, viral load, and age.
We fuse the ACTG~175 trial \citep{hammer1996} with the Multicenter AIDS Cohort Study \citep{kaslow1987}.
The cohort followed participants over more than 25 years with periodic visits.

The fusion of a trial with real-world data poses three challenges for survival outcomes.
The first is unmeasured confounding in the real-world data.
Treatment is not randomised, and assignment may depend on factors the measured covariates do not fully capture.
The no-unmeasured-confounders assumption is untestable and often violated in practice \citep{vanderweele2017evalue}.
The second is heterogeneity between the two data sources.
Even when the treatment effect transports across sources, the baseline prognosis often does not.
The trial and observational populations typically differ in eligibility, follow-up, and case mix.
A naive analysis confounds this baseline difference with the treatment effect.
This heterogeneity is not confined to the baseline prognoses: differences in measurement protocol and follow-up quality also reshape the error distribution across sources.
The third is censoring.
Right censoring arises when the event has not occurred by the end of follow-up, so the time to event is known only to exceed the censoring time.
The real-world data may also be interval-censored, with events recorded only at periodic visits.

The Bayesian fusion forest addresses the three challenges in a single nonparametric model.
For the first challenge, a confounding function absorbs the unmeasured confounding in the real-world data.
For the second challenge, a shared baseline prognosis with a source-specific deviation captures the heterogeneity between the sources, and a source-specific error distribution captures heterogeneity beyond the mean.
For the third challenge, an accelerated failure time formulation accommodates right- and interval-censored event times.
Existing methods meet some of these challenges but none meets all three.
We will now provide more context on the challenges and solutions.

For the first challenge, the confounding function measures the confounding bias in the real-world data.
It was introduced as a sensitivity-analysis device within a single study \citep{robins2000sensitivity}.
The confounding function can be estimated by combining a randomised and an observational study.
\citet{Kallus2018} learn a parametric confounding function from the RCT and debias the observational treatment effect.
\citet{YangLiuZengWang2025} develop a semiparametric-efficient estimator that jointly estimates the treatment effect and the confounding function, and derive a test for unmeasured confounding.
The elastic integrative estimator of \citet{yang2023elastic} uses that test to decide whether to borrow from the real-world data, and discards it when confounding is detected.
Two frequentist methods extend the approach to survival outcomes.
\citet{Ye2025} work under an accelerated failure time model with right censoring and high-dimensional covariates, and treat the confounding function as a sparse component selected jointly with the treatment effect.
\citet{mao2025statistical} target the conditional restricted mean survival time under right censoring.
Neither accommodates interval censoring, the third challenge.
Here, we extend the confounding function framework to an accelerated failure time model with interval censoring, so that challenges one and three are both addressed.
We target the acceleration factor: a causally interpretable estimand on the survival-time scale.

Bayesian methods address the second challenge of between-source heterogeneity by borrowing information through the prior or the likelihood.
The power prior of \citet{ibrahim2000power} raises the external likelihood to a fractional power that controls the borrowing.
The meta-analytic-predictive prior of \citet{neuenschwander2010} makes the borrowing hierarchical and adaptive to between-source heterogeneity.
Later variants tie the borrowing to the between-source discrepancy through commensurate priors \citep{hobbs2012commensurate}.
These methods all target a marginal or control-arm effect.
More recent literature targets heterogeneous effects directly.
\citet{zhou2021incorporating} pool trial and external outcome surfaces with Bayesian additive regression trees, but do not allow for unmeasured confounding in the external source.
\citet{dimitriou2025causal} develop a multi-task Gaussian process with a data-adaptive borrowing parameter.
None of these Bayesian methods targets heterogeneous treatment effects on survival outcomes under unmeasured confounding.
We encode the borrowing in the priors: a mean-zero deviation centres the prior on equality of the two baseline prognoses, and the confounding function is shrunk towards a covariate-independent bias.

We impose a Bayesian additive regression tree (BART) prior \citep{chipman2010bart} on each component of our model, a popular tool in causal machine learning \citep{hill2011bayesian}.
BART has theoretical guarantees \citep{rockova2020posterior} and strong empirical performance in both estimation accuracy and uncertainty quantification \citep{dorie2019, thal2023, kabata2026}.
The uncertainty quantification is intrinsic to the model and follows directly from the posterior distribution.
BART captures nonlinear and interaction effects without requiring us to specify them.
The regularisation of BART is governed by a small number of interpretable hyperparameters, which lets us tune the prior of each model component to its substantive role.
The error distribution deserves the same flexibility: it plausibly differs between the two data sources, but the two laws are unlikely to be unrelated.
We pair the trees with a nonparametric error distribution: a hierarchical Dirichlet process mixture lets the error law differ between the sources in shape and scale while sharing mixture components.
Machine-learning predictors such as (deep) neural networks or gradient-boosted trees offer comparable flexibility, but they target prediction rather than causal effect estimation and provide no mechanism for data fusion.
We benchmark our framework against four such methods in the simulation study, and improve on all of them in both accuracy and calibration.

\section{Methodology}

\subsection{Causal framework}

We consider data from two sources: a randomised controlled trial (RCT) and a real-world data study (RWD).
The data-source indicator $S \in \{0, 1\}$ takes the value $S = 1$ for RCT observations and $S = 0$ for RWD observations.
The RCT contributes $n_1$ independent observations and the RWD contributes $n_0$ independent observations.
We observe a vector of pre-treatment covariates $X \in \mathbb{R}^p$ and a binary treatment $A \in \{0, 1\}$ for each individual.
We denote the random variables without the subject index $i$ in general.
The outcome of interest is the nonnegative survival time $T$, subject to censoring.
We allow two types of censoring: right censoring, and interval censoring.
For each subject, we observe a pair $(L, R)$ with $0 \leq L \leq R \leq \infty$ satisfying $T \in [L, R]$, together with a censoring-type indicator $\delta \in \{0, 1, 2\}$.
The indicator $\delta$ distinguishes three observation types, each defined by a relation between the pair $(L, R)$ and the survival time $T$.
For exact observation ($\delta = 1$), $L = R = T$, so the survival time is observed directly.
For right censoring ($\delta = 0$), $L = C$ and $R = \infty$ for a censoring time $C$, so $T$ is known only to exceed $C$.
For interval censoring ($\delta = 2$), $0 \leq L < R < \infty$, so $T$ is known only to lie in the bounded interval $[L, R]$.
The standard right-censored setup is recovered when $\delta \in \{0, 1\}$ for every subject.
The two sources may carry different censoring types: in our application the trial is right-censored while the cohort is interval-censored.
We write $\mathcal{C}$ for the inspection and censoring process underlying $(L, R, \delta)$, and refer to \citet{sun2006statistical} for details.
We adopt the potential-outcomes framework \citep{rubin1974}.
For $a \in \{0, 1\}$, the potential survival time $T(a)$, the censoring process $\mathcal{C}(a)$, and the observation $(L(a), R(a), \delta(a))$ are the quantities that would be observed if treatment were set to $a$.

We estimate the conditional average treatment effect (CATE):
\begin{equation}
\tau(x) = \E[\log T(1) - \log T(0) \mid X = x].
\label{eq:cate}
\end{equation}
The CATE describes how the treatment effect varies across covariates.
The corresponding estimand on the multiplicative time scale is the \emph{acceleration factor} $\exp\{\tau(x)\}$ \citep{pang2021flexible}.
We choose this estimand for its interpretation and its causal properties.
The acceleration factor admits a direct interpretation on the survival-time scale: at every quantile of the survival distribution, treatment rescales the survival time by a constant multiplicative factor.
This time-scaling is more intuitive in a clinical setting than the relative change in event rate conveyed by a hazard ratio \citep{Swindell2009}.
The acceleration factor also enjoys causal properties that the hazard ratio lacks.
\citet{Brathovde2024} show that the observed acceleration factor identifies its causal counterpart under exchangeability and consistency.
This identification continues to hold under unmeasured frailty and treatment-effect heterogeneity.
The hazard ratio loses its causal interpretation in those same settings, because conditioning on survival induces a built-in selection bias \citep{hernan2010}.
Moreover, the acceleration factor is collapsible: the marginal and conditional acceleration factors coincide when the treatment effect is homogeneous, regardless of the distribution of unmeasured baseline risk \citep{Crowther2023}.

We impose the following assumptions on the causal structure and the censoring mechanism.
\vspace*{-\parskip}
\begin{adjustwidth}{1.5em}{1.5em}
\setlength{\parskip}{0pt}
\begin{assumption}[Consistency]\label{ass:sutva}
If $A = a$ then $T = T(a)$ and $\mathcal{C} = \mathcal{C}(a)$ for each $a \in \{0, 1\}$.
\end{assumption}

\begin{assumption}[Unconfoundedness of the RCT]\label{ass:rct-unconf}
$T(a) \,\ind\, A \mid X,\, S = 1$ for each $a\in\{0, 1\}$.
\end{assumption}

\begin{assumption}[Positivity]\label{ass:positivity-both}
$0 < \p(A = 1 \mid X = x,\, S = s) < 1$ for each $s \in \{0, 1\}$ and all $x$ in the support of $X \mid S = s$.
\end{assumption}

\begin{assumption}[Cross-source transportability]\label{ass:transport}
$\E[\log T(1) - \log T(0) \mid X,\, S = 1] \;=\; \E[\log T(1) - \log T(0) \mid X,\, S = 0]$.
\end{assumption}

\begin{assumption}[Conditionally non-informative censoring]\label{ass:censoring}
$T(a) \,\ind\, \mathcal{C}(a) \mid X,\, A,\, S$ for each $a\in\{0, 1\}$.
\end{assumption}

\end{adjustwidth}
\setlength{\parskip}{0.8em}
\vspace{0.8em}

Assumptions~\ref{ass:sutva}, \ref{ass:positivity-both} and~\ref{ass:censoring} are standard in causal inference.
Consistency requires a well-defined treatment and no interference between subjects.
Together with the construction of $(L, R, \delta)$ from the survival time and the censoring process, consistency extends to the observed data: on $\{A = a\}$ the observed triple equals its potential counterpart.
Positivity requires each covariate profile to have a positive probability of either treatment within its source.
It holds by design in the trial and is an overlap condition on the real-world data \citep{dahabreh2019generalizing}.
Assumption~\ref{ass:censoring} is the conditional version of non-informative censoring standard in survival analysis \citep{sun2006statistical}.
The censoring and inspection processes may depend on covariates, treatment and source but not on the survival time beyond these.
Assumptions~\ref{ass:rct-unconf} and~\ref{ass:transport} carry the substantive content of our approach.

Our identification strategy relaxes unconfoundedness in the RWD and retains transportability of the CATE across sources.
Assumption~\ref{ass:rct-unconf} holds by design in a randomised trial.
Treatment in the RWD may depend on confounders not contained in $X$.
The resulting bias is absorbed by the confounding function $c(x)$ introduced in the next section.
We assume the CATE to be time-invariant by not explicitly letting it depend on time.
Other approaches that combine evidence from multiple sources take the opposite route: they allow source-specific CATEs but require unconfoundedness within each source \citep{stuart2011propensity, dahabreh2019generalizing, shyr2025multistudy}.
The two relaxations cannot coexist in our setting.
Without unconfoundedness of the RWD, the CATE and $c(x)$ are not separately identifiable unless the CATE transports across sources.
We therefore allow for confounding in the real-world data at the explicit cost of retaining Assumption~\ref{ass:transport}.
We make this interplay explicit in Propositions~\ref{prop:decomp} and~\ref{prop:ident} in the next section.

\subsection{Accelerated failure time decomposition}

We model the conditional log survival time jointly across the trial and the real-world data.
In the real-world data, the treated-versus-control contrast generally differs from the causal treatment effect.
Treatment assignment may depend on confounders outside $X$, which bias the observed contrast.
We define the \emph{confounding function} $c(x)$ \citep{robins2000sensitivity} as:
\begin{equation}
c(x) \;=\; \E[\log T \mid X = x,\, A = 1,\, S = 0] \,-\, \E[\log T \mid X = x,\, A = 0,\, S = 0] \,-\, \tau(x).
\label{eq:cf}
\end{equation}
The confounding function captures the difference between the RWD treated-versus-control contrast and the true causal effect.
The source indicator $S$ ties the two sources into one model.
Intuitively, a baseline prognosis and the causal effect enter in both sources, while the confounding bias enters only in the real-world data.
Proposition~\ref{prop:decomp} makes this intuition precise.
Proposition~\ref{prop:ident} shows how the decomposition translates to identification results.
\begin{proposition}[Accelerated failure time decomposition]\label{prop:decomp}
Under Assumptions~\ref{ass:sutva}--\ref{ass:transport}, the conditional log survival time satisfies:
\begin{equation}
\E[\log T \mid A,\, X,\, S] \;=\; m_0(X, S) \,+\, \tau(X)\,A \,+\, (1 - S)\,A\,c(X),
\label{eq:decomp}
\end{equation}
where $m_0(X, S) := \E[\log T \mid A = 0,\, X,\, S]$ is the baseline log survival time.
\end{proposition}

The baseline prognosis $m_0(X, S)$ is the survival a patient would have without treatment, which can differ between the two sources.
The confounding term carries the factor $(1 - S)$, so it acts only in the real-world data.

\begin{proposition}[Identification]\label{prop:ident}
Under Assumptions~\ref{ass:sutva}--\ref{ass:transport}:
\begin{enumerate}[label=\textup{(\roman*)}]
\item The CATE is identified from the RCT alone:
\begin{equation}
\tau(x) \;=\; \E[\log T \mid X = x,\, A = 1,\, S = 1] \,-\, \E[\log T \mid X = x,\, A = 0,\, S = 1].
\label{eq:ident-tau}
\end{equation}
\item Given $\tau$, the confounding function is identified from the RWD.
\item Neither $\tau$ nor $c$ is identified from the RWD alone; the RWD identifies only the composite $\tau(x) + c(x)$.
\end{enumerate}
\end{proposition}

Proofs of Propositions~\ref{prop:decomp} and~\ref{prop:ident} are in Supplementary Materials~\ref{sm:proofs}.
The same identification results hold for the acceleration factor $\exp\{\tau(x)\}$ by continuity of the exponential.
Similar identification results appear in \citet{YangLiuZengWang2025} and \citet{Ye2025} for a propensity-residualised contrast.
Propositions~\ref{prop:decomp} and~\ref{prop:ident} concern the population conditional means $\E[\log T \mid A, X, S]$ of the log transformed survival time.
Assumptions~\ref{ass:sutva}--\ref{ass:transport} identify the causal quantities from these conditional means.
Assumption~\ref{ass:censoring} on the censoring mechanism plays a separate role: it lets us recover the conditional means from the censored observations $(L, R, \delta)$.
The censoring assumption enters the likelihood and the data augmentation in the sampler, not the causal identification argument.
This separation enables us to handle interval censoring in addition to right censoring in the sequel.
The non-identification in part~(iii) is structural rather than statistical.
Larger real-world samples cannot resolve it.
Separating $\tau$ from $c$ requires external information that identifies one of the two functions.
The trial supplies it through part~(i).

We model the log survival time with the accelerated failure time (AFT) specification implied by the decomposition:
\begin{equation}
\log T \;=\; m_0(X, S) \,+\, \tau(X)\,A  \,+\, (1 - S)\,A\,c(X) \,+\, \varepsilon.
\label{eq:aft-model}
\end{equation}
We assume the errors are independent across subjects and have mean zero.
We impose no parametric form on their distribution and allow it to differ across sources through a hierarchical prior.
The regression functions $m_0$, $\tau$ and $c$ retain the causal meanings established above.
We assign each function a separate Bayesian tree ensemble prior.

\subsection{Nonparametric error distribution}

Four requirements guide our model for the error $\varepsilon$ in \eqref{eq:aft-model}.
The first requirement is flexibility in shape.
Log survival times often show skewness or multimodality.
Censoring makes the fit particularly sensitive to the assumed shape.
An overly restrictive error model may induce bias in the estimates.
The second requirement is that the error distribution may differ between the sources.
Randomised trials and real-world data sources typically differ in measurement protocol, follow-up quality and the prevalence of extreme outcomes.
These differences plausibly enter the error distribution rather than the structural mean.
The third requirement is that the two source-specific distributions remain linked.
They are unlikely to be entirely unrelated.
A model that treats them as fully independent forfeits useful pooling.
The loss is greatest when one source is small.
The fourth requirement is a mean-zero error in each source so that $m_0$, $\tau$ and $c$ keep their conditional-mean interpretation.
Existing data-fusion methods typically assume a single shared error law for both sources, often a Gaussian.
Such a law fails the first two requirements.
We meet all four with a hierarchical Dirichlet process mixture of normals (HDPM).
The prior is source-specific: $\varepsilon \mid S = s \sim F_s$, where $F_s$ is the distribution of the error term in source $s$.
We let $F_0$ and $F_1$ differ flexibly while sharing a common set of mixture components across sources.

Conditional on the source indicator $S = s$, we model the error as a location mixture of Gaussians with source-specific scale $\sigma_s > 0$ and mixing measure $G_s$:
\begin{equation}
\varepsilon \mid S = s,\, G_s,\, \sigma_s \;\sim\; \int \frac{1}{\sigma_s}\, \phi\!\left(\frac{w - \theta}{\sigma_s}\right) dG_s(\theta).
\label{eq:hdp-fs}
\end{equation}
Here $\phi$ is the standard normal density and $\theta$ is the location of a mixture component.
The right-hand side gives the density of $F_s$, evaluated at the argument $w$.
Gaussian-mixture priors of this form have been used for AFT residuals in single-source BART models \citep{henderson2020individualized}.
The mixing measures share a latent structure through the hierarchical Dirichlet process \citep{teh2006hierarchical} with a top-level random measure $G^*$:
\begin{equation}
G^* \;\sim\; \mathrm{DP}(\gamma, H), \qquad G_s^* \mid G^*,\, M_s \;\sim\; \mathrm{DP}(M_s, G^*), \qquad s \in \{0, 1\},
\label{eq:hdp-prior}
\end{equation}
with base distribution $H = \mathcal{N}(0, \sigma_\theta^2)$, top-level concentration parameter $\gamma$, and source-specific concentration parameters $M_s$.
The top-level measure is discrete and admits a stick-breaking representation:
\begin{equation}
G^* \;=\; \sum_{k=1}^\infty \beta_k\, \delta_{\theta_k^*}, \qquad \theta_k^* \stackrel{\mathrm{iid}}{\sim} H, \qquad \beta_k = u_k \prod_{l < k}(1 - u_l), \qquad u_k \stackrel{\mathrm{iid}}{\sim} \mathrm{Beta}(1, \gamma).
\label{eq:hdp-stick}
\end{equation}
Each source-specific measure $G_s^*$ is supported on the same atoms $\{\theta_k^*\}_{k\geq 1}$ as $G^*$, with its own weight vector $\bm{\pi}_s = (\pi_{sk})_{k \ge 1}$.
The shape of each error distribution $F_s$ is therefore parameterised by a shared set of mixture components and a source-specific weighting over them.

It remains to enforce the fourth requirement of a mean-zero error in each source.
We follow \citet{yang2010semiparametric} and obtain centred mixing measures by subtracting the source-specific mean of the unconstrained measure:
\begin{equation}
\mu_s \;=\; \int \theta\, dG_s^*(\theta) \;=\; \sum_{k=1}^\infty \pi_{sk}\, \theta_k^*, \qquad G_s \;=\; \sum_{k=1}^\infty \pi_{sk}\, \delta_{\theta_k^* - \mu_s}.
\label{eq:hdp-centring}
\end{equation}
By construction $\E[\varepsilon \mid S = s,\, G_s,\, \sigma_s] = 0$.
The unconstrained atoms $\theta_k^*$ remain shared across sources, while the centred atoms $\theta_k^* - \mu_s$ are source-specific through the shifts $\mu_0$ and $\mu_1$.
We place a scaled-inverse-$\chi^2$ prior $\sigma_s^2 \sim \nu\lambda / \chi^2_\nu$ on the per-source residual scales, with degrees of freedom $\nu$ and scale $\lambda$.
We set $\nu = 3$ following the BART default and calibrate $\lambda$ to the empirical residual variance \citep{chipman2010bart}.

\subsection{Overview of BART}

We build each regression function in our model on Bayesian additive regression trees (BART) and adopt the standard formulation of \citet{chipman2010bart}.
Three properties motivate this choice.
BART captures nonlinear and interaction effects without requiring us to specify them.
Its regularisation is governed by a small number of interpretable hyperparameters, which lets us tune the prior of each component to its substantive role in the next section.
Its conjugate structure yields efficient Gibbs updates that extend to censored outcomes.
BART places a flexible nonparametric prior over an unknown function $f : \mathbb{R}^p \to \mathbb{R}$.
The prior represents $f$ as a sum of $J$ regression trees:
\begin{equation}
f(x) \;=\;
  \sum_{j=1}^{J} g(x;\, \mathcal{T}_j, \mathcal{H}_j),
\label{eq:bart-revised}
\end{equation}
where each $\mathcal{T}_j$ is a binary tree with $b_j$ terminal nodes.
Each interior splitting rule has the form $\{x_\ell \leq x^*\}$ for some covariate index $\ell$ and splitting value $x^*$.
We write $\mathcal{H}_j = \{h_{j,1}, \ldots, h_{j,b_j}\}$ for the step heights at those terminal nodes.
The function $g(x;\, \mathcal{T}_j, \mathcal{H}_j)$ routes $x$ through the splits of $\mathcal{T}_j$ to a terminal node $r$ and returns the step height $h_{j,r}$ at that node.

The prior on $(\mathcal{T}_1, \mathcal{H}_1), \ldots, (\mathcal{T}_J, \mathcal{H}_J)$ is independent across trees.
It decomposes into a prior on the tree-structure $\mathcal{T}_j$ and a prior on the step heights $\mathcal{H}_j$ given the tree structure $\mathcal{T}_j$.
The structure of each $\mathcal{T}_j$ follows the recursive splitting process of \citet{chipman1998bayesian}.
A node at depth $d$ is non-terminal with probability $\alpha(1+d)^{-\beta}$.
The hyperparameters $\alpha$ and $\beta$ regularise tree depth.
Conditional on the tree, the splitting variable $x_\ell$ at each interior node is drawn uniformly from the available covariates.
The splitting value $x^*$ is drawn uniformly from its observed values.
The step heights receive independent conjugate normal priors $h_{j,r} \mid \mathcal{T}_j \sim \mathcal{N}(0,\, \sigma_h^2)$ with $\sigma_h = k/(2\sqrt{J})$ and $k > 0$ the step-height scale.
A horseshoe prior on the step heights is a more suitable alternative in high-dimensional settings \citep{jacobs2025horseshoe}.
\citet{chipman2010bart} centre and rescale the response, then choose $k$ so that the implied prior on $f$ assigns circa $95\%$ of its mass to the observed response range.
We denote the BART prior with these hyperparameters by $\mathrm{BART}(J, k, \alpha, \beta)$.
The four arguments are the number of trees $J$, the step-height scale $k$, and the tree-structure hyperparameters $\alpha$ and $\beta$.

\subsection{The Bayesian fusion forest}

We place a separate Bayesian additive regression tree prior on each component of the decomposition~\eqref{eq:decomp}: the baseline prognosis $m_0$, the treatment effect $\tau$, and the confounding function $c$.
Each component carries an independent causal meaning (Proposition~\ref{prop:decomp}), so each prior acts directly on an interpretable quantity.
An undifferentiated BART on the full conditional mean $\E[\log T \mid A, X, S]$ would reach the priors on $\tau$ and $c$ only indirectly.
The induced prior on a contrast then depends on incidental features such as the dimension and distribution of $X$ \citep{hahn2020bayesian}.
It also cannot encode the between-source borrowing and the confounding shrinkage that the problem calls for.
The decomposition sacrifices nothing in expressiveness, since any conditional mean of the form~\eqref{eq:decomp} stays in the support.
It only changes which functions the prior deems likely.
This separate-forest construction generalises the prognostic-treatment split of the Bayesian causal forest \citep{hahn2020bayesian} to two data sources and an explicit confounding function.
We tune each prior through its tree-structure hyperparameters and its ensemble size.
We order the regularisation along a ladder, from the flexible baseline prognosis to the strongly regularised confounding function.

Two considerations order the ladder: the information the data carry about each regression function and the complexity we expect of that function.
The baseline prognosis enters the mean of every observation and may be complex, so it receives the most flexible prior.
The treatment effect is identified from the trial contrast alone (Proposition~\ref{prop:ident}) and typically varies with few effect modifiers, so we shrink it towards a homogeneous effect \citep{hahn2020bayesian}.
This shrinkage also reduces the risk of spurious effect heterogeneity.
The confounding function is estimated only from the difference between the real-world contrast and the treatment effect.
The confounding bias stems from covariates we do not observe, so we shrink the confounding function towards uniformity.

The baseline prognosis $m_0(X, S)$ is the first component, and it may differ between the two sources.
We split it into a shared component and a source-specific deviation:
\begin{equation}
m_0(X, S) \;=\; m_0^{\mathrm{sh}}(X) + (1 - S)\, d(X),
\label{eq:m0-decomp-prop}
\end{equation}
where $m_0^{\mathrm{sh}}$ is shared across both sources and the deviation $d$ is active only in the real-world data.
The deviation $d$ is the difference between the two source-specific baselines.
We place a separate BART prior on $m_0^{\mathrm{sh}}$ and on $d$, and we give $d$ a mean-zero prior.
The prior then centres on equality of the two source baselines, which encodes a preference for borrowing.
A single BART on $m_0(X, S)$ with $S$ as a covariate cannot encode this preference.
Two independent baselines would imply a needlessly diffuse prior on their difference.
The shared-plus-deviation form instead places the prior directly on the between-source difference and tunes the borrowing through the deviation scale $k_d$.
We read this split as a meta-analytic-predictive prior on the prognostic function \citep{neuenschwander2010}.
A small $k_d$ shrinks the deviation towards zero for strong borrowing, and a large $k_d$ lets the real-world baseline depart (Supplementary Materials~\ref{sm:map-m0}).

We let $m_0^{\mathrm{sh}}$ be the most flexible forest in the model.
We use the tree-structure hyperparameters $(\alpha_{\mathrm{sh}}, \beta_{\mathrm{sh}}) = (0.95, 2)$ and $J_{\mathrm{sh}} = 200$ trees, so the shared baseline can accommodate complex disease-intrinsic structure.
We use the same $(\alpha_d, \beta_d) = (0.95, 2)$ for $d$ but only $J_d = 50$ trees.
The smaller ensemble suffices because $d$ only absorbs residual differences in protocol and patient population.
The deviation keeps the same permissive tree-structure prior, so it can track covariate-dependent differences between the sources.

The treatment effect $\tau$ is the second component, and we regularise it more strongly than the baseline.
We place a depth-penalised BART prior on $\tau$ to discourage higher-order interactions in $X$.
We set $\beta_\tau = 3$ and keep $\alpha_\tau = 0.95$, so the prior mass concentrates on trees of depth one or two.
We reduce the ensemble to $J_\tau = 100$ trees, which further strengthens the regularisation.
At the no-split extreme the treatment forest reduces to a constant, a homogeneous treatment effect.
The prior thus shrinks the conditional average treatment effect towards homogeneity.

The confounding function $c$ is the third component, and we regularise it most strongly of all.
We regularise it through both the root-split probability and the depth penalty.
We set the base splitting probability $\alpha_c = 0.25$ and the depth parameter $\beta_c = 3$, and use $J_c = 50$ trees.
The low base probability discourages splits at the root.
The depth penalty discourages deep interactions when splits do occur.
The settings $(\alpha_c, \beta_c) = (0.25, 3)$ express a strong prior preference for confounding that varies little with the covariates, close to a uniform bias.

We standardise the log survival time before fitting by the centring and scaling constants of a preliminary log-normal AFT fit.
This rescales the response to a unit-variance scale.
Each forest carries its own prior step-height scale $k_f$, indexed by $f \in \{\mathrm{sh}, d, \tau, c\}$ for the shared baseline $m_0^{\mathrm{sh}}$, the deviation $d$, the treatment effect $\tau$, and the confounding function $c$.
The step-height scales calibrate the prior magnitude of each ensemble to the scale of the standardised response.
We set $k_f = 1$ for every forest by default.
We halve it for the treatment effect, $k_\tau = 1/2$, which shrinks the magnitude of $\tau$ on top of the structural regularisation already placed on it.
We tune further by cross-validation.

We summarise the full hierarchical model, which we refer to as the \emph{Bayesian fusion forest}:
\begin{align*}
\log T_i &\;=\; m_0^{\mathrm{sh}}(X_i) + (1 - S_i)\, d(X_i) + A_i\, \tau(X_i) + (1 - S_i)\, A_i\, c(X_i) + \varepsilon_i, \\
m_0^{\mathrm{sh}} &\;\sim\; \mathrm{BART}(200,\, k_{\mathrm{sh}},\, 0.95,\, 2), \\
d &\;\sim\; \mathrm{BART}(50,\, k_d,\, 0.95,\, 2), \\
\tau &\;\sim\; \mathrm{BART}(100,\, k_\tau,\, 0.95,\, 3), \\
c &\;\sim\; \mathrm{BART}(50,\, k_c,\, 0.25,\, 3), \\
\varepsilon_i \mid S_i = s,\, G_s,\, \sigma_s &\;\sim\; \int \sigma_s^{-1}\, \phi\!\left(\tfrac{w - \theta}{\sigma_s}\right) dG_s(\theta), \\
G_s &\;=\; \textstyle\sum_k \pi_{sk}\, \delta_{\theta_k^* - \mu_s}, \quad s \in \{0, 1\}, \\
G_s^* \mid G^*,\, M_s &\;\sim\; \mathrm{DP}(M_s,\, G^*), \quad s \in \{0, 1\}, \\
G^* \mid \gamma &\;\sim\; \mathrm{DP}(\gamma,\, H), \qquad H = \mathcal{N}(0,\, \sigma_\theta^2), \\
\gamma &\;\sim\; \mathrm{Gamma}(a_\gamma,\, b_\gamma), \\
M_s &\;\sim\; \mathrm{Gamma}(a_M,\, b_M), \quad s \in \{0, 1\}, \\
\sigma_s^2 &\;\sim\; \nu\lambda / \chi^2_\nu, \quad s \in \{0, 1\}.
\end{align*}

We calibrate the base-measure variance $\sigma_\theta^2$ rather than assign it a prior.
Supplementary Materials~\ref{sm:posterior-computation} give the hyperparameter values.

\subsection{Posterior inference}

We sample from the posterior via a blocked Gibbs sampler.
We update the BART forests $m_0^{\mathrm{sh}}$, $d$, $\tau$, and $c$ in turn against the partial residual.
The partial residual subtracts the other forests and the current draw of the residual mixture from the augmented outcome.
Each update reduces to a standard BART update on a Gaussian working response \citep{chipman2010bart}.
We augment the censored observations \citep{tanner1987}.
The HDPM error block is updated by a separate Gibbs step over its cluster assignments, atoms, mixture weights, and concentration parameters.
We give the full sampler in the Supplementary Materials.

We compute average treatment effects by marginalising the conditional effect $\tau(\cdot)$ against a target covariate distribution.
We take the target to be the union of the two source studies, with covariate distribution $F_X = \pi_0 F_X^0 + \pi_1 F_X^1$.
Here $F_X^s$ is the covariate distribution within source $s \in \{0, 1\}$ and $\pi = (\pi_0, \pi_1)$ lies on the unit simplex.
The posterior of the marginal effect carries uncertainty from three sources: the posterior over $\tau$, the within-source covariate distributions $F_X^s$, and the mixing fractions $\pi$.
We propagate all three jointly by a hierarchical Bayesian bootstrap.
The choice of $\pi$ encodes the analyst's target population.
Natural deterministic choices include the empirical fractions $\pi_s = n_s / n$ and equal weighting $\pi = (1/2, 1/2)$.
The source-specific limits $\pi = (1, 0)$ and $\pi = (0, 1)$ target the population of the RWD and RCT respectively.
We may instead place a Dirichlet prior $\pi \sim \mathrm{Dirichlet}(\alpha)$ on the mixing fractions, with concentration parameter $\alpha = (\alpha_0, \alpha_1)$.
This propagates uncertainty about the target composition.
The choice $\alpha = (n_0, n_1)$ centres the prior on the empirical fractions and approximately recovers a pooled Bayesian bootstrap.
We recommend this as the default.
The choice $\alpha = (1, 1)$ gives a flat prior on the simplex.
We draw mixing fractions $\pi^{(b)} \sim \mathrm{Dirichlet}(\alpha)$ for each posterior draw $\tau^{(b)}(\cdot)$ of the conditional effect.
We draw within-source weights $w^{(s, b)} \sim \mathrm{Dirichlet}(\mathbf{1}_{n_s})$ over the observations of source $s$ independently for $s \in \{0, 1\}$.
The corresponding draw of the average treatment effect is:
\begin{equation}
\bar{\tau}^{(b)} \;=\; \sum_{s \in \{0, 1\}} \pi_s^{(b)} \sum_{i: S_i = s} w_i^{(s, b)}\, \tau^{(b)}(X_i).
\label{eq:ate-bb}
\end{equation}
The within-source weights are the Bayesian bootstrap \citep{Rubin1981} applied to each $F_X^s$.
 
We can summarise each fitted forest by projecting it on a lower-dimensional model.
The posterior draws of $m_0^{\mathrm{sh}}$, $d$, $\tau$, and $c$ retain the causal meaning established in Proposition~\ref{prop:decomp}.
We can therefore summarise any one of them in isolation.
We project a fitted forest onto a simple second-stage model since a full posterior over a nonparametric surface is hard to communicate.
This follows the posterior-summarisation approach of \citet{Woody2020} and the two-stage logic of the Virtual Twins method \citep{foster2011subgroup}.
The first stage is the fitted forest.
The second is a low-complexity model fit to the posterior-mean surface of that forest, such as a single regression tree or a linear projection.
We propagate posterior uncertainty by fixing the second-stage structure at its posterior-mean fit and recomputing its parameters on every posterior draw.

\section{Simulation study}
We compare the Bayesian fusion forest on the combined data sources with an accelerated failure time Bayesian causal forest \citep{jacobs_software} on the RCT and RWD alone.
The causal forest is the single-source counterpart of the Bayesian fusion forest, so the comparison isolates the value of combining the sources rather than differences in model class.
The trial-only fit is unbiased but limited by the trial's size.
The real-world-only fit uses the larger sample but is exposed to confounding.
The Bayesian fusion forest must improve on both to justify combining the sources.
We assess its robustness to unmeasured confounding and to between-source heterogeneity, benchmark it against flexible machine-learning methods, and study its behaviour as the covariate dimension grows.

There are currently very few methods to reasonably compare the Bayesian fusion forest to.
In particular, we found no data fusion method for right- and interval-censored outcomes.
The closest alternatives are flexible machine-learning methods for survival prediction.
We compare with four such methods: an accelerated failure time deep neural network \citep{norman2024deepaft}, gradient boosting under an accelerated failure time loss \citep{barnwal2022xgboost}, and Buckley--James boosting over extreme learning machines and over regression trees \citep{kong2023bjelm}.
These methods share three structural limitations.
The first limitation is that they target prediction rather than causal effect estimation.
The second limitation is that they provide no mechanism for combining a randomised and a confounded source.
The third limitation is that most of them lack support for interval censoring.
We make several adaptations to render the methods suitable for comparison with the Bayesian fusion forest.
We embed each method in an S- and a T-learner to obtain treatment-effect estimates \citep{kunzel2019metalearners}.
We fit each learner on the trial alone and on a naive pool of both sources.

\subsection{Simulation setup}

The general setup is common to all three experiments below.
The trial contributes $n_1 = 150$ observations and the real-world data $n_0 = 350$.
We draw $p = 10$ covariates $X \sim \mathcal{N}(0, \Sigma)$ in both sources, with an AR(1) correlation $\Sigma_{ij} = \rho^{|i-j|}$, $\rho = 0.3$.
Five covariates are active and five are noise.
We generate the outcome:
\begin{equation}
  \begin{aligned}
    \log T_{\RCT} &= m_0(x) + A\, \tau(x) + \varepsilon_{\RCT}, \\
    \log T_{\RWD} &= m_0(x) + \lambda_d\, d(x) + A\, \tau(x)
                     + \lambda_u\, A\, U + \varepsilon_{\RWD},
  \end{aligned}
  \label{eq:sim-dgp}
\end{equation}
with components:
\begin{equation}
  m_0(x) = 2 x_1 - x_2 x_3 + \tfrac{1}{2} x_4^2, \quad \tau(x) = \tfrac{1}{2} + x_1 - \tfrac{1}{2} x_2^2, \quad d(x) = x_4 - \tfrac{1}{2} x_5.
\end{equation}
The function $\tau$ is the true CATE on the log-time scale.
We randomise treatment in the trial, $A \sim \mathrm{Bernoulli}(1/2)$, and let real-world treatment depend on an unmeasured confounder $U \sim \mathrm{Uniform}(0, 1)$ through $A \sim \mathrm{Bernoulli}\bigl\{\expit(x_1 + U)\bigr\}$.
The confounder enters the outcome only in the treated arm.
The factor $\lambda_d$ controls the between-source heterogeneity in baseline prognosis and $\lambda_u$ the strength of the unmeasured confounding.
We model the trial error as $\varepsilon_{\RCT} \sim \mathcal{N}(0, \sigma^2)$ with $\sigma = 3/4$, and draw the real-world error from a standardised Gumbel law: $\varepsilon_{\RWD} = \tfrac{\sqrt{6}}{\pi}\, \gamma_{\mathrm{E}} - G$ with $G \sim \mathrm{Gumbel}(0, \sqrt{6}/\pi)$, which has mean zero, unit variance, and is skewed.
Trial and real-world event times are conditionally log-normal and Weibull, respectively.
We right-censor the trial by an exponential censoring time tuned to circa $35\%$ censoring.
We interval-censor the real-world data at eight inspection times.
Events after the last inspection are right-censored, circa $20\%$ of the real-world observations.
The two censoring types mirror the application.

We fit the Bayesian fusion forest to the combined data with default parameters.
We run 1000 Monte Carlo replications.
Each fit draws 5000 posterior samples after 5000 burn-in iterations.
We report the root mean squared error, bias, $95\%$ credible-interval coverage, and posterior variance of the estimated CATE.

We investigate the performance of the Bayesian fusion forest in three separate simulation setups.
\begin{enumerate}[label=\arabic*), leftmargin=*]
  \item \textbf{Unmeasured confounding and between-source heterogeneity.}
    We investigate how the bias and RMSE of the Bayesian fusion forest behave as the unmeasured confounding and the between-source heterogeneity grow.
    We vary the confounding strength $\lambda_u \in \{0, 0.5, 1, 1.5, 2\}$ at fixed $\lambda_d = 1$, and the heterogeneity $\lambda_d \in \{0, 0.5, 1, 1.5, 2\}$ at fixed $\lambda_u = 1$.
  \item \textbf{Comparison with machine-learning methods.}
    We compare the Bayesian fusion forest with the four machine-learning methods in each of the four learner--source configurations.
    We tune each method by cross-validation and obtain confidence intervals from 100 bootstrap resamples.
    We fix $\lambda_d = \lambda_u = 1$.
    Not all methods handle interval censoring, so for this experiment we right-censor the real-world data at circa $35\%$ and every method sees the same data.
  \item \textbf{Increasing covariate dimension.}
    We investigate how the precision gained by borrowing from the real-world data trades off against the cost of searching a larger covariate space, and whether fusion remains worthwhile in higher covariate space dimensions.
    We hold the sample sizes fixed and grow the number of covariates from $p = 5$ to $p = 500$ at $\lambda_d = \lambda_u = 1$.
    Only five covariates carry signal, so the remaining covariates are pure noise and the problem grows sparser.
    We draw the covariates in independent blocks of ten with the AR(1) correlation $\rho = 0.3$ within each block.
    We also fit an oracle variant given only the five active covariates.
    It marks the best performance attainable when the active covariates are known.
\end{enumerate}

We additionally test the gain from modelling the error distribution nonparametrically.
We compare the hierarchical Dirichlet process mixture with a single Gaussian and a pooled Dirichlet process mixture, both of which force the two sources to share one error law.
The nonparametric model is robust across error laws and improves on both alternatives (Supplementary Materials~\ref{sm:sim-additional}).

\subsection{Simulation results}

\begin{figure}[!tb]
  \centering
  \includegraphics[width=0.8\textwidth]{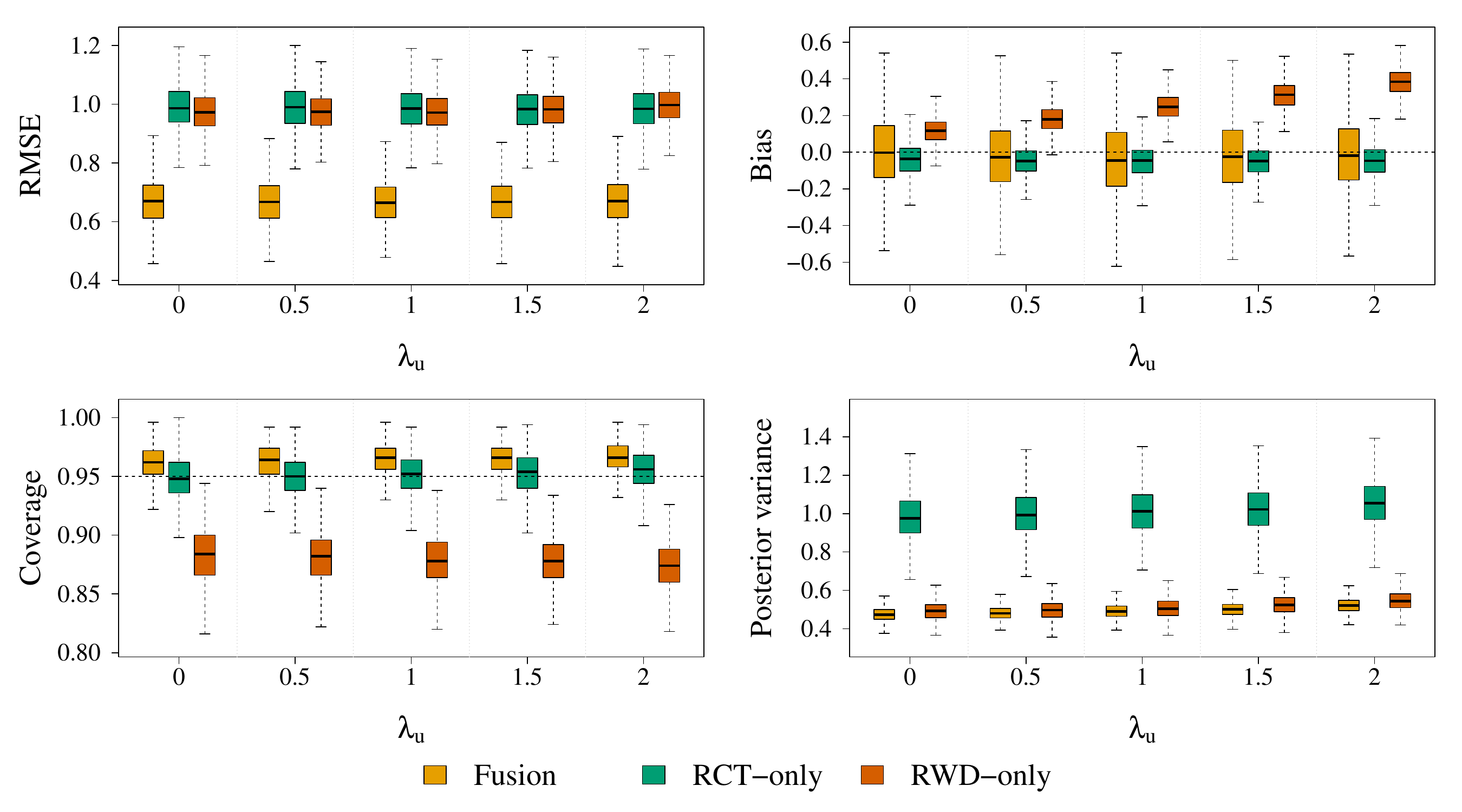}
  \caption{CATE metrics under varying unmeasured confounding strength $\lambda_u$, with the RWD prognostic deviation held at $\lambda_d = 1$. From top-left to bottom-right: pointwise RMSE, mean bias, 95\% credible-interval coverage, and posterior variance of $\hat\tau(x)$, averaged over the evaluation set and across simulation replicates. Dashed reference lines indicate zero bias and nominal coverage.}
  \label{fig:sim-lambda-u}
\end{figure}

We first vary the confounding strength $\lambda_u$ and hold the between-source heterogeneity fixed at $\lambda_d = 1$.
Figure~\ref{fig:sim-lambda-u} reports the results.
The real-world-only estimate is unbiased at $\lambda_u = 0$.
Its bias then grows roughly linearly with $\lambda_u$, to circa $0.4$ at $\lambda_u = 2$.
The trial-only estimate is unaffected by $\lambda_u$.
The Bayesian fusion forest stays essentially unbiased across the whole range.
The Bayesian fusion forest also attains a lower RMSE than the trial-only baseline at every level.
Coverage is near nominal for the fusion and trial-only estimates.
The real-world-only estimate loses coverage as its bias grows.
The fusion posterior variance is roughly half that of the trial-only baseline throughout.
This reflects the precision gained by borrowing from the larger real-world sample.

We report the results for varying between-source heterogeneity $\lambda_d$ at fixed confounding $\lambda_u = 1$ in Supplementary Materials~\ref{sm:sim-additional}.
All three estimators are essentially flat in $\lambda_d$.
The fusion and trial-only estimates stay unbiased and at nominal coverage.
The real-world-only estimate keeps the bias from the fixed confounding.
RMSE and posterior variance are stable across $\lambda_d$.

We compare the Bayesian fusion forest with the machine-learning methods at $\lambda_d = \lambda_u = 1$.
Table~\ref{tab:sim-competitors} reports all four methods in all four configurations.
The Bayesian fusion forest attains the lowest root mean squared error of every method and configuration, and it is the only method whose intervals reach nominal coverage at a competitive width.
The alternatives that cover do so with intervals between $1.2$ and $2.1$ times as wide, and those with narrower intervals cover between $0.00$ and $0.48$.
Pooling the two sources without a confounding function transfers the confounding bias: within every family the pooled fits shift the bias upward relative to their trial-only counterparts.
The deep neural network attains the lowest error of the alternatives in three of the four configurations.
It is therefore the strongest of the four, and we compare against it alone in the remainder of the paper.
The S-learners show undercoverage, a consequence of their narrow bootstrap intervals.
The T-learners recover coverage, but only by widening their intervals, and the pooled deep network overcovers outright.
We give the full results across the $(\lambda_d, \lambda_u)$ grid in Supplementary Materials~\ref{sm:sim-additional}.

\begin{table}[!tb]
  \centering
  \caption{Comparison with the machine-learning
    alternatives at $\lambda_d = \lambda_u = 1$: a deep neural network (DNN),
    gradient boosting under an accelerated failure time loss (XGBoost), and
    Buckley--James boosting over extreme learning machines (BJ-ELM) and
    regression trees (BJ-trees), each fitted as an S- and a T-learner on the
    trial alone and on a naive pool of both sources. Both sources are
    right-censored at circa $35\%$. Coverage and width refer to $95\%$
    intervals: posterior for the Bayesian fusion forest and percentile
    bootstrap over $100$ resamples otherwise. Averages over $1000$
    replications. The lowest root mean squared error and the bias closest to
    zero are in bold. The BJ-trees S-learner never selects
    the treatment indicator, so its estimated effect and intervals are
    identically zero.}
  \label{tab:sim-competitors}
  \small
  \begin{tabular}{@{}lll rrrr@{}}
    \toprule
    Method & Source & Learner & RMSE & Bias & Coverage & Width \\
    \midrule
    Bayesian fusion forest & Both & --- & $\mathbf{0.676}$ & $\mathbf{-0.026}$ & 0.964 & 2.82 \\
    \midrule
    DNN                    & Trial & S & 1.115 & $-0.185$ & 0.375 & 0.95 \\
                           &       & T & 1.250 & $-0.247$ & 0.913 & 3.73 \\
                           & Pool  & S & 1.107 & 0.105 & 0.444 & 0.99 \\
                           &       & T & 1.350 & 0.137 & 0.991 & 4.92 \\
    \addlinespace
    XGBoost                & Trial & S & 1.161 & $-0.038$ & 0.158 & 0.40 \\
                           &       & T & 1.593 & $-0.075$ & 0.937 & 4.95 \\
                           & Pool  & S & 1.114 & 0.151 & 0.269 & 0.64 \\
                           &       & T & 1.120 & 0.306 & 0.905 & 3.27 \\
    \addlinespace
    BJ-ELM                 & Trial & S & 1.281 & $-0.068$ & 0.445 & 1.42 \\
                           &       & T & 1.662 & 0.393 & 0.952 & 5.97 \\
                           & Pool  & S & 1.282 & 0.309 & 0.300 & 0.90 \\
                           &       & T & 1.175 & 0.472 & 0.896 & 3.34 \\
    \addlinespace
    BJ-trees               & Trial & S & 1.209 & $-0.040$ & 0.000 & 0.00 \\
                           &       & T & 1.354 & $-0.054$ & 0.460 & 1.73 \\
                           & Pool  & S & 1.222 & 0.028 & 0.000 & 0.00 \\
                           &       & T & 1.013 & 0.343 & 0.475 & 1.25 \\
    \bottomrule
  \end{tabular}
\end{table}

\begin{figure}[!tb]
  \centering
  \includegraphics[width=0.9\textwidth]{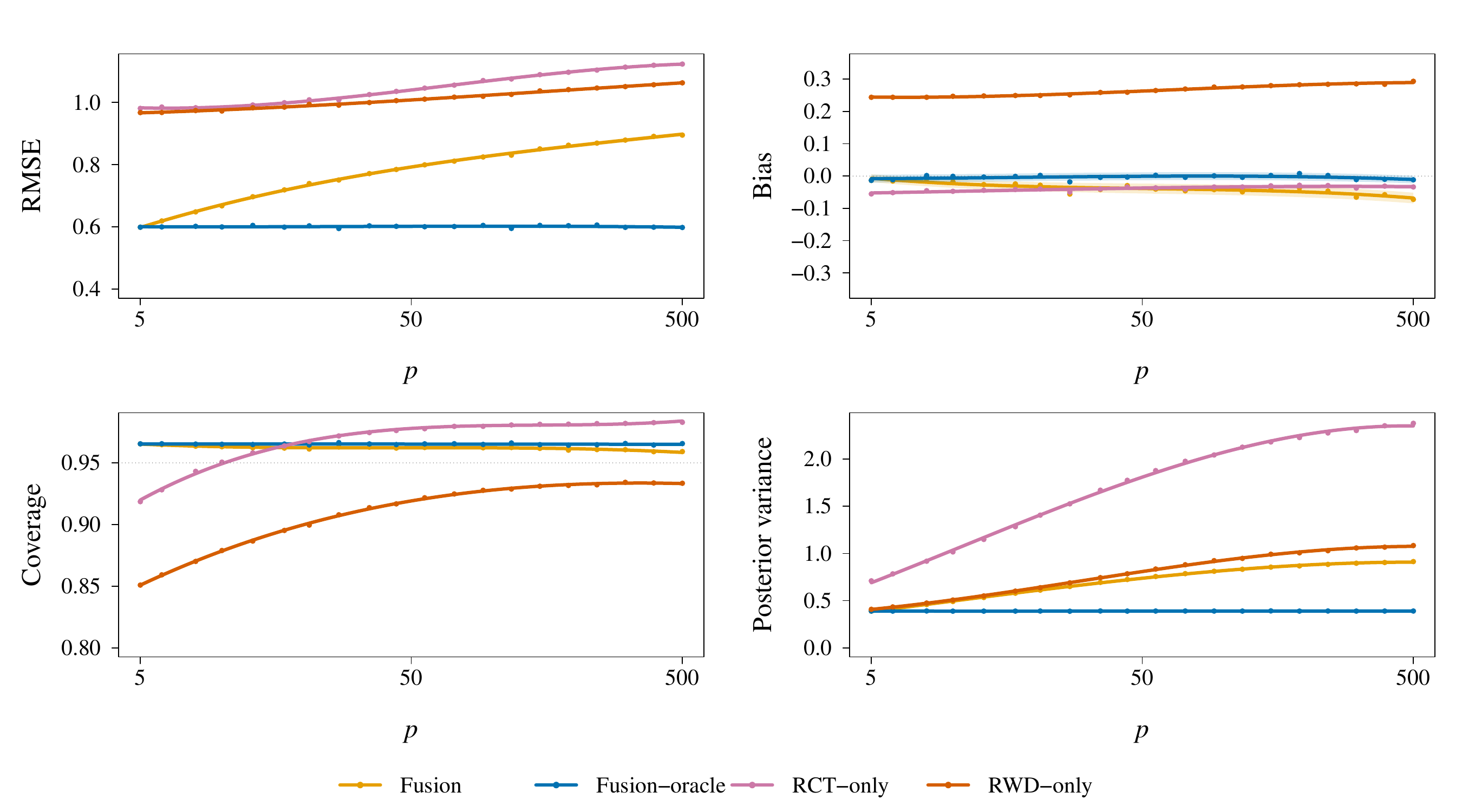}
  \caption{CATE metrics as the number of covariates $p$ grows from $5$ to $500$, with $n_1 = 150$, $n_0 = 350$, and $\lambda_d = \lambda_u = 1$; five covariates are active and the remaining $p - 5$ are noise. From top-left to bottom-right: RMSE, bias, $95\%$ credible-interval coverage, and posterior variance of $\hat\tau(x)$. The fusion oracle is the Bayesian fusion forest given only the five active covariates, and traces the achievable floor. Dashed reference lines mark zero bias and nominal coverage.}
  \label{fig:sim-highdim}
\end{figure}

We finally examine robustness to the covariate dimension.
Figure~\ref{fig:sim-highdim} reports the CATE metrics against $p$.
The Bayesian fusion forest stays essentially unbiased across the whole range, as does the trial-only baseline, while the real-world-only estimate retains its confounding bias.
All estimators lose precision as $p$ grows, but at very different rates.
The trial-only posterior variance increases several-fold, whereas the fusion variance grows only mildly, and the fusion attains the lowest RMSE of the feasible estimators at every $p$.
The fusion holds near-nominal coverage throughout, while the trial-only intervals turn conservative as their variance increases.
The efficiency of the fusion relative to the trial-only baseline confirms that the gain persists.
The RMSE ratio rises from $0.68$ at $p = 10$ to $0.81$ at $p = 500$, yet stays below one throughout.
The posterior-variance ratio stays near one-half (Supplementary Materials~\ref{sm:sim-additional}).
Because both estimators are unbiased, this advantage is essentially a variance effect.
The fusion borrows the larger real-world sample to control a variance the trial alone cannot.

These results show that the Bayesian fusion forest is robust to both nuisances.
The Bayesian fusion forest corrects for the unmeasured confounding in the real-world data and stays unbiased as the confounding strengthens.
It accommodates the prognostic heterogeneity between the sources and is unaffected as the heterogeneity grows.
It also exploits the larger real-world sample and achieves lower RMSE and posterior variance than the trial-only baseline at every grid point.

\section{Application: survival after HIV}

\begin{figure}[!tb]
  \centering
  \includegraphics[width=0.8\linewidth]{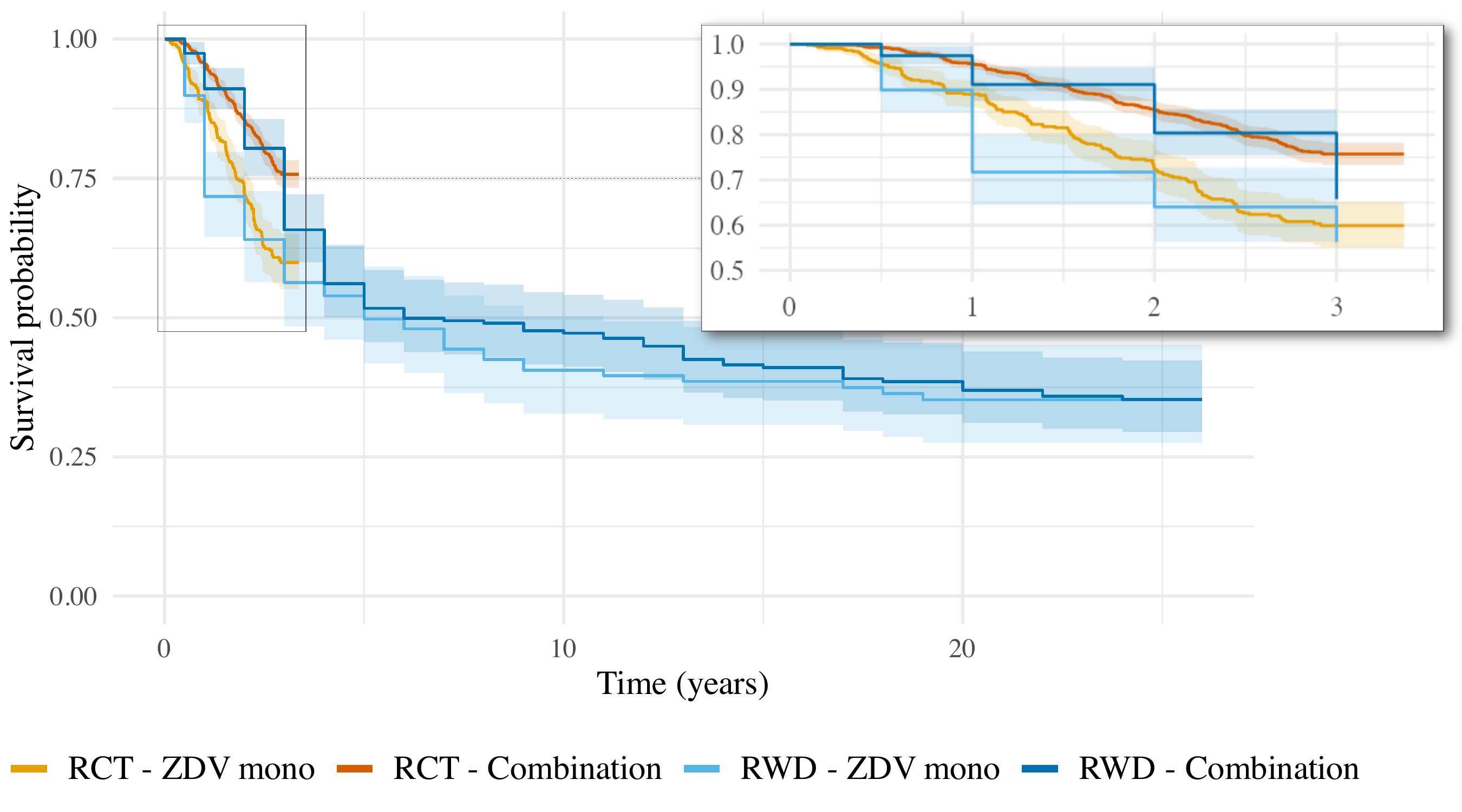}
  \caption{Kaplan--Meier curves of event-free survival, by data source
    (RCT, RWD) and treatment group (ZDV monotherapy and combination).
    The main panel spans the full cohort follow-up of circa 25 years;
    the inset zooms to the shorter trial window. The real-world data curves are
    approximate due to interval censoring.}
  \label{fig:km-curves}
\end{figure}

We study the effect of a combination antiretroviral therapy versus zidovudine (ZDV) monotherapy on event-free survival in HIV.
We combine the AIDS Clinical Trials Group Protocol~175 (ACTG~175) \citep{hammer1996} with the Multicenter AIDS Cohort Study (MACS) \citep{kaslow1987}.
Treatment in ACTG~175 is randomised.
We pool the three combination regimens in ACTG~175 into one comparator and contrast it with ZDV monotherapy.
We restrict both sources to a common population: men with baseline CD4 count of 200 to 500 cells/mm$^3$.
This alignment makes the fusion credible, since the two sources then approximate one population.
We adjust for six baseline covariates: age, CD4, CD8, calendar year, race, and years of prior antiretroviral therapy.
We summarise the aligned cohort in Supplementary Materials~\ref{sm:analysis-additional}.
The aligned cohort contains 1771 trial patients and 373 real-world data patients.
The trial follows patients for a maximum of circa three years: too short to reach median event-free survival.
The MACS study follows them for circa 25 years and reaches a median event-free survival of five to six years.
This longer follow-up adds robustness for the late treatment effect where the trial is uninformative.
MACS records outcomes at annual visits.
The data give event times only to the calendar year, so the events are interval-censored at annual resolution.
Figure~\ref{fig:km-curves} shows the Kaplan--Meier survival curves.

We fit a Bayesian fusion forest to the combined data with default parameter settings.
We compare it with an AFT Bayesian causal forest fitted to the trial alone.
We compute the average effect with a source-specific Bayesian bootstrap.
The source weights follow a Dirichlet prior proportional to the sample sizes.
The fusion estimate of the average acceleration factor is $1.65$ ($95\%$ credible interval $[1.43;1.90]$).
The trial-only baseline is close at $1.70$ ($95\%$ credible interval $[1.51;1.92]$).
Combination therapy thus stretches the event-free survival time scale by a factor of $1.65$ relative to monotherapy.
Equivalently, the median event-free survival time under combination therapy is $1.65$ times the median under monotherapy.

\begin{figure}[!tb]
  \centering
  \includegraphics[width=0.8\linewidth]{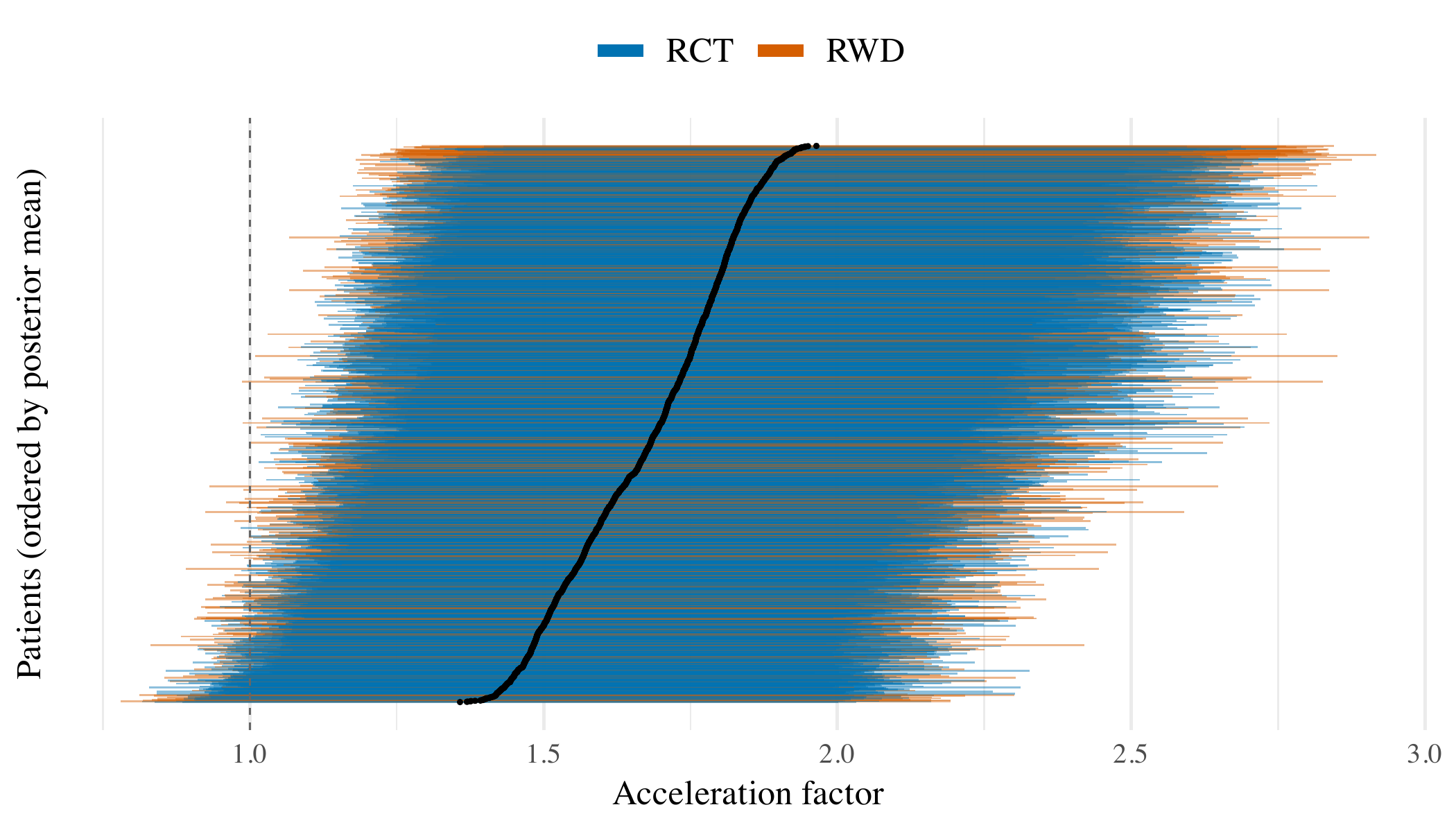}
  \caption{Subject-level acceleration factor for every
    patient in the trial-aligned cohort, from the Bayesian fusion forest,
    ordered by posterior mean. Vertical bars are pointwise 95\%
    credible intervals; colour indicates the data source (RCT, RWD);
    the dashed line at $1$ marks no effect.}
  \label{fig:cate}
\end{figure}

The Bayesian fusion forest estimates each patient's effect with lower posterior variance than the trial alone.
Figure~\ref{fig:cate} shows the subject-level acceleration factor for every patient, ordered by posterior mean.
The estimates lie above one, so combination helps almost everyone.
The size of the gain differs across patients.
The fusion intervals are far narrower than the trial-only intervals.
The average 95\% credible-interval width falls from $3.16$ to $1.21$ on the acceleration-factor scale, a reduction of $62\%$ (Supplementary Materials~\ref{sm:analysis-additional}).
The fusion sharpens the individual estimates.
This contrasts with the average effects presented above: the fusion interval there is slightly wider, because the Bayesian bootstrap propagates uncertainty about the target population's covariate distribution.

We also fit the flexible machine-learning predictors of the simulation study to this cohort.
None of them targets a causal contrast, and none offers a mechanism for combining a randomised with a confounded source.
We fit a deep accelerated failure time network, gradient-boosted trees, and Buckley--James boosting over extreme learning machines and over regression trees, each as an S- and a T-learner, on the trial alone and on a naive pool of both sources.
Across the sixteen resulting configurations the share of patients declared at least $95\%$ certain to benefit ranges from $0.0\%$ to $100\%$, and no principled rule selects among them.
Interval width cannot decide between the methods here, because the cohort supplies no ground truth: a narrow interval may reflect genuine precision or mere overconfidence.
We report the comparison in full in Supplementary Materials~\ref{sm:analysis-competitors}.
The Bayesian fusion forest instead gives a single coherent answer, with calibrated intervals and a posterior probability of benefit for every patient.

We investigate which patient characteristics drive the variation in benefit.
We summarise the treatment-effect forest $\tau$ by the posterior projection, with a single regression tree as the second-stage model (CART; \citealp{breiman1984cart}).
Figure~\ref{fig:fit-the-fit} shows a depth-three tree on the six covariates.
It splits on CD8 count and race.
The other covariates do not enter.
The acceleration factor rises with CD8, from circa $1.5$ in the lowest group to circa $1.8$ in the highest.
White patients gain slightly more than non-white within each CD8 band.
Every leaf exceeds $1.4$, so all subgroups benefit.
The treatment benefit is largest at high CD8.

\definecolor{splitfill}{RGB}{230,244,250}
\begin{figure}[!tb]
  \centering
  \resizebox{0.8\textwidth}{!}{%
  \begin{tikzpicture}[
    every node/.style = {align=center, font=\footnotesize},
    splitn/.style     = {draw, rounded corners=2pt, fill=splitfill,
                         minimum width=1.15cm, minimum height=0.55cm,
                         inner sep=2.5pt},
    leafn/.style      = {draw, rounded corners=2pt, fill=white,
                         minimum width=1.45cm, minimum height=0.85cm,
                         inner sep=3pt, font=\scriptsize},
    elabel/.style     = {midway, fill=white, inner sep=4pt,
                         font=\scriptsize},
    edgeln/.style     = {draw=black!55, thick}]

    \node[splitn] (root)  at ( 0.35, 4.7) {CD8};
    \node[splitn] (mid1)  at (-3.00, 3.4) {CD8};
    \node[splitn] (mid2)  at ( 3.70, 3.4) {CD8};
    \node[splitn] (raceA) at (-4.50, 1.9) {race};
    \node[splitn] (raceB) at ( 1.70, 1.9) {race};
    \node[splitn] (raceC) at ( 5.70, 1.9) {race};

    \node[leafn] (L1) at (-5.4, 0) {$1.45$\\$[1.05;1.92]$\\$n = 124$};
    \node[leafn] (L2) at (-3.6, 0) {$1.51$\\$[1.19;1.87]$\\$n = 381$};
    \node[leafn] (L3) at (-1.4, 0) {$1.60$\\$[1.31;1.92]$\\$n = 330$};
    \node[leafn] (L4) at ( 0.8, 0) {$1.65$\\$[1.25;2.13]$\\$n = 107$};
    \node[leafn] (L5) at ( 2.6, 0) {$1.72$\\$[1.43;2.07]$\\$n = 425$};
    \node[leafn] (L6) at ( 4.8, 0) {$1.76$\\$[1.33;2.32]$\\$n = 178$};
    \node[leafn] (L7) at ( 6.6, 0) {$1.83$\\$[1.51;2.22]$\\$n = 599$};

    \draw[edgeln] (root) -- (mid1)  node[elabel] {$\le 800$};
    \draw[edgeln] (root) -- (mid2)  node[elabel] {$> 800$};

    \draw[edgeln] (mid1) -- (raceA) node[elabel] {$\le 650$};
    \draw[edgeln] (mid1) -- (L3)    node[elabel] {$> 650$};
    \draw[edgeln] (mid2) -- (raceB) node[elabel] {$\le 1065$};
    \draw[edgeln] (mid2) -- (raceC) node[elabel] {$> 1065$};

    \draw[edgeln] (raceA) -- (L1) node[elabel] {non-white};
    \draw[edgeln] (raceA) -- (L2) node[elabel] {white};
    \draw[edgeln] (raceB) -- (L4) node[elabel] {non-white};
    \draw[edgeln] (raceB) -- (L5) node[elabel] {white};
    \draw[edgeln] (raceC) -- (L6) node[elabel] {non-white};
    \draw[edgeln] (raceC) -- (L7) node[elabel] {white};
  \end{tikzpicture}}
  \caption{Regression-tree summary of the acceleration factor
    $\exp\{\tau(x)\}$ from the Bayesian fusion forest. We fit the tree to the
    posterior-mean log-time CATE on the six baseline covariates. Each leaf
    gives the within-leaf average acceleration factor, as a posterior mean
    and 95\% credible interval.}
  \label{fig:fit-the-fit}
\end{figure}
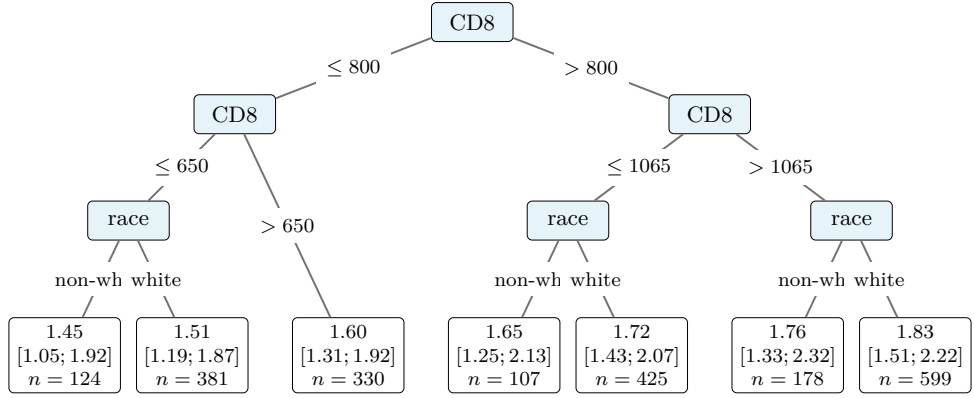

We inspect the posterior probability of treatment benefit $\p(\tau(x_i) > 0 \mid \text{data})$ for each patient \citep{henderson2020individualized}.
We bin the patients by posterior probability of benefit and report the fraction in each bin.
We compute it over the trial population, under the fusion and trial-only fits, so the columns are comparable (Table~\ref{tab:hte-summary}).
We find strong evidence of benefit for almost every patient.
The trial alone is far less certain, with only $38\%$ above $0.95$ and a quarter of patients inconclusive.
The Bayesian fusion forest gains this certainty at the conditional level even though its average effect is slightly smaller.

We investigate the between-source heterogeneity in baseline prognosis.
We summarise the deviation forest $d(x)$ by a projection on a linear model.
The deviation forest measures how the baseline prognosis differs between the trial and the real-world data.
We project $d(\cdot)$ onto the covariates by least squares and present the results in Table~\ref{tab:projection}.
The gap is dominated by a constant shift.
The intercept is $1.70$ ($95\%$ credible interval $[1.31;2.10]$), so the observational baseline sits well above the trial.
Among the covariates, only prior antiretroviral years moves the gap, by $0.27$ ($95\%$ credible interval $[-0.01;0.55]$).
Age, CD4, CD8, calendar year, and race have coefficients near zero, with intervals covering zero.
The baseline prognosis therefore differs mainly through a large average shift.
This shift likely reflects population and measurement differences which the six covariates do not capture.

Combination therapy prolongs event-free survival by a factor of circa $1.65$.
We identify benefit for nearly every patient, far more than the trial alone.
This near-universal advantage echoes earlier ACTG~175 analyses.
The original trial \citep{hammer1996} and later re-analyses on survival-type endpoints \citep{lee2026bjboost} all favour combination therapy over monotherapy.
The Buckley--James Q-learner \citep{lee2026bjboost} likewise recommends it for over $96\%$ of participants.
We found that the treatment effect varies with CD8 count and race.
This agrees with evidence that the benefit of combination therapy varies with baseline risk \citep{kennedy2023}.
The baseline prognosis differs between the sources, mainly through a large average shift and in part through its dependence on prior antiretroviral exposure.
The fusion delivers this certainty with calibrated uncertainty.

\begin{table}[!tb]
  \centering
  \caption{Posterior probability of benefit over the trial
    population. Higher bins give stronger evidence of benefit, the
    lowest evidence of harm. Entries are fractions of patients.}
  \label{tab:hte-summary}
  \small
  \begin{tabular}{@{}lrr@{}}
    \toprule
    $\p(\tau(x_i) > 0 \mid \text{data})$        & Fusion & Trial-only \\
    \midrule
    $(0.99, 1]$    & $0.767$ & $0.180$ \\
    $(0.95, 0.99]$ & $0.206$ & $0.198$ \\
    $(0.75, 0.95]$ & $0.028$ & $0.364$ \\
    $(0.25, 0.75]$ & $0.000$ & $0.245$ \\
    $[0, 0.25]$    & $0.000$ & $0.014$ \\
    \bottomrule
  \end{tabular}
\end{table}

\begin{table}[!tb]
  \centering
  \caption{Linear projection of the deviation forest
    $d(x)$ onto the covariates: posterior mean coefficient and $95\%$
    credible interval on the log-transformed time scale. The CD4 and CD8
    coefficients are reported per $100$ cells/mm$^3$.}
  \label{tab:projection}
  \small
  \begin{tabular}{@{}lrr@{}}
    \toprule
    Term          & Coefficient & $95\%$ credible interval \\
    \midrule
    Intercept     & $1.701$  & $[1.312;2.099]$ \\
    \midrule
    Age, y                    & $-0.003$ & $[-0.020;0.015]$ \\
    CD4, per 100 cells/mm$^3$ & $0.061$  & $[-0.064;0.189]$ \\
    CD8, per 100 cells/mm$^3$ & $0.005$  & $[-0.019;0.032]$ \\
    Calendar year & $-0.032$ & $[-0.256;0.194]$ \\
    Race          & $0.071$  & $[-0.304;0.443]$ \\
    Prior ART, y  & $0.269$  & $[-0.008;0.550]$ \\
    \bottomrule
  \end{tabular}
\end{table}

\section{Discussion}

We combined a randomised trial with real-world data to estimate heterogeneous treatment effects on survival outcomes, without assuming the real-world source is unconfounded.
The simulation study and the HIV application demonstrate its advantages over a single-source analysis.
The Bayesian fusion forest remains unbiased under varying levels of unmeasured confounding.
It reduces the posterior variance of a trial-only analysis.
In the application, we recovered a benefit of combination therapy that the survival curves of the real-world data alone did not reveal.
We established benefit for nearly every patient where the trial alone was inconclusive.
The model handled the right-censored trial and the interval-censored cohort jointly.

Our framework implicitly assumes the acceleration factor $\exp\{\tau(X)\}$ is time-invariant: treatment rescales the survival time by a factor that depends on the covariates but not on time.
This assumption may be inappropriate when treatment effects unfold over time, such as delayed-onset benefits or waning effects.
Flexible parametric AFT models relax this assumption by letting the acceleration factor vary with time \citep{Crowther2023}.
The causal interpretation of time-dependent acceleration factors has been formalised \citep{Brathovde2024}.
A natural extension replaces $\tau(X)$ with a generalised Bayesian tree ensemble $\tau(X, t)$.

\section*{Data availability}
\sloppy
The software implementation is available on GitHub at \url{https://github.com/tijn-jacobs/FusionForests}.
The same repository also holds the code for the simulation studies and the data analysis.
A release on the Comprehensive R Archive Network is planned.
The ACTG~175 trial data are available in the \texttt{speff2trial} R package \citep{speff2trial}.
The MACS data can be requested through the MWCCS cohort website.

\section*{Funding}
This project has received funding from the European Research Council (ERC)
under the European Union's Horizon Europe program under Grant agreement
No.~101074802.
Views and opinions expressed are however those of the author(s) only and do not
necessarily reflect those of the European Union or the European Research
Council Executive Agency.
Neither the European Union nor the granting authority can be held responsible
for them.
This work used the Dutch national e-infrastructure with the support of the SURF
Cooperative using grant no.~EINF-18803.

\section*{Acknowledgements}
The authors used generative artificial intelligence tools in support of
developing and debugging the accompanying software.
All such output was reviewed and verified by the authors, who take full
responsibility for the content of this work.

\section*{Conflict of interest}
None declared.

\FloatBarrier
\bibliographystyle{apalike}
\bibliography{references}


\startsupplement

\begin{center}
  {\Large\bfseries Supplementary Materials}
\end{center}

\medskip


\section{Proofs}\label{sm:proofs}
\subsection{Proof of Proposition~\ref{prop:decomp}}
We show:
\begin{equation}
\E[\log T \mid A, X, S] \;=\; m_0(X, S) \,+\, \tau(X)\,A \,+\, (1 - S)\,A\,c(X),
\label{eq:sm-decomp}
\end{equation}
at each of the four values of $(A, S)$:
The left-hand side is the conditional mean of the observed $\log T$.
The right-hand side decomposes that mean into causal quantities: a source-specific baseline $m_0(X, S)$, the treatment effect $\tau(X)$, and the confounding bias $c(X)$.

We use two elementary arguments throughout the proof.
By consistency (Assumption~\ref{ass:sutva}), we replace the observed outcome by the relevant potential outcome on the event where treatment is fixed: $\log T = \log T(a)$ on $\{A = a\}$.
By RCT unconfoundedness (Assumption~\ref{ass:rct-unconf}), we have $T(a) \,\ind\, A \mid X, S = 1$.
Inside the RCT subpopulation $\{S = 1\}$, the conditional distribution of $T(a)$ given $X$ is unchanged by further conditioning on $\{A = a\}$.
The two conditional expectations coincide: $\E[\log T(a) \mid A = a, X, S = 1] = \E[\log T(a) \mid X, S = 1]$.
Together, the two arguments yield, for each $a \in \{0, 1\}$:
\begin{equation}
\E[\log T \mid A = a,\, X,\, S = 1] \;=\; \E[\log T(a) \mid X,\, S = 1].
\label{eq:po-step}
\end{equation}

\emph{Case $A = 0$, $S \in \{0, 1\}$.}
The two treated-arm terms in \eqref{eq:sm-decomp}, $\tau(X)\,A$ and $(1 - S)\,A\,c(X)$, vanish.
The right-hand side reduces to $m_0(X, S)$.
The left-hand side equals $m_0(X, S)$ by the definition $m_0(X, S) := \E[\log T \mid A = 0, X, S]$.

\emph{Case $A = 1$, $S = 1$ (RCT).}
We apply \eqref{eq:po-step} at $a = 1$ to obtain:
\begin{equation}
\E[\log T \mid A = 1,\, X,\, S = 1] \;=\; \E[\log T(1) \mid X,\, S = 1].
\end{equation}
We split $\log T(1) = \log T(0) + [\log T(1) - \log T(0)]$ inside the expectation:
\begin{equation}
\E[\log T(1) \mid X,\, S = 1] \;=\; \E[\log T(0) \mid X,\, S = 1] \,+\, \E[\log T(1) - \log T(0) \mid X,\, S = 1].
\label{eq:po-split}
\end{equation}
The first term equals $m_0(X, 1)$ by \eqref{eq:po-step} at $a = 0$:
\begin{equation}
\E[\log T(0) \mid X,\, S = 1] \;=\; \E[\log T \mid A = 0,\, X,\, S = 1] \;=\; m_0(X, 1).
\end{equation}
By cross-source transportability (Assumption~\ref{ass:transport}), the conditional log-treatment effect does not depend on $S$.
The contrast on the right of \eqref{eq:po-split} therefore equals the population CATE $\tau(X) = \E[\log T(1) - \log T(0) \mid X]$ defined in the manuscript:
\begin{equation}
\E[\log T(1) - \log T(0) \mid X,\, S = 1] \;=\; \E[\log T(1) - \log T(0) \mid X] \;=\; \tau(X).
\end{equation}
We combine the three results: $\E[\log T \mid A = 1, X, S = 1] = m_0(X, 1) + \tau(X)$.
This matches the right-hand side of \eqref{eq:sm-decomp} at $(A, S) = (1, 1)$.
The confounding-function term $(1 - S)\,A\,c(X)$ vanishes when $S = 1$.

\emph{Case $A = 1$, $S = 0$ (RWD).}
We recall the definition of the confounding function from the manuscript:
\begin{equation}
c(x) \;=\; \E[\log T \mid X = x,\, A = 1,\, S = 0] \,-\, \E[\log T \mid X = x,\, A = 0,\, S = 0] \,-\, \tau(x).
\label{eq:sm-cf}
\end{equation}
We rearrange:
\begin{equation}
\E[\log T \mid A = 1,\, X,\, S = 0] \;=\; m_0(X, 0) + \tau(X) + c(X),
\end{equation}
which equals the right-hand side of \eqref{eq:sm-decomp} at $(A, S) = (1, 0)$.

This case uses none of the manuscript's causal-framework assumptions directly.
They enter only through $\tau$: they identify $\tau$ as the causal CATE rather than an unrelated function of $X$.
That identification is the content of Proposition~\ref{prop:ident}(i), proved below.\qed

\subsection{Proof of Proposition~\ref{prop:ident}}
We use positivity (Assumption~\ref{ass:positivity-both} of the manuscript) throughout the proof.
It guarantees that the conditional expectations below are well-defined for almost every $x$ in the relevant support.

\emph{Part (i): identification of $\tau$ from the RCT.}
We evaluate \eqref{eq:sm-decomp} at the two values of $A$ within the RCT slice $\{S = 1\}$.
The confounding-function term $(1 - S)\,A\,c(X)$ vanishes when $S = 1$:
\begin{align}
\E[\log T \mid A = 1,\, X,\, S = 1] &\;=\; m_0(X, 1) + \tau(X),\\
\E[\log T \mid A = 0,\, X,\, S = 1] &\;=\; m_0(X, 1).
\end{align}
The difference of the two conditional means cancels $m_0(X, 1)$ and leaves $\tau(X)$:
\begin{equation}
\tau(X) \;=\; \E[\log T \mid A = 1,\, X,\, S = 1] \,-\, \E[\log T \mid A = 0,\, X,\, S = 1].
\label{eq:ident-tau-pf}
\end{equation}
Both conditional means on the right are functionals of the joint distribution of the observed outcomes $(\log T, A, X) \mid S = 1$.
This identifies $\tau$ from the RCT alone.

The identification rests on two assumptions.
By RCT unconfoundedness (Assumption~\ref{ass:rct-unconf} of the manuscript), the observed RCT contrast at $X = x$ equals the conditional treatment effect within the RCT population.
By cross-source transportability (Assumption~\ref{ass:transport} of the manuscript), this RCT-population effect coincides with the population CATE.
Neither the RWD data nor the confounding function $c$ enters \eqref{eq:ident-tau-pf}.

\emph{Part (ii): identification of $c$ from the RWD, given $\tau$.}
The confounding function is defined in \eqref{eq:sm-cf} as:
\begin{equation}
c(X) \;=\; \E[\log T \mid A = 1,\, X,\, S = 0] \,-\, \E[\log T \mid A = 0,\, X,\, S = 0] \,-\, \tau(X).
\label{eq:ident-c-pf}
\end{equation}
The first two terms are functionals of the RWD data distribution.
If $\tau$ is already identified, then \eqref{eq:ident-c-pf} identifies $c$.
Part (i) supplies such a $\tau$ from the RCT.
This is the identification strategy underlying the confounding-function approach to combining randomised and observational data \citep{Kallus2018, YangLiuZengWang2025}.

\emph{Part (iii): the RWD alone does not identify $\tau$ and $c$ separately.}
We show that distinct parameter pairs $(\tau, c)$ produce the same RWD distribution.
The decomposition \eqref{eq:sm-decomp} restricted to $\{S = 0\}$ yields two equations in three unknown functions $m_0(\cdot, 0)$, $\tau$, and $c$:
\begin{align}
\E[\log T \mid A = 0,\, X,\, S = 0] &\;=\; m_0(X, 0), \label{eq:rwd-control}\\
\E[\log T \mid A = 1,\, X,\, S = 0] &\;=\; m_0(X, 0) + \tau(X) + c(X). \label{eq:rwd-treated}
\end{align}
The first equation identifies $m_0(\cdot, 0)$.
The second equation identifies the composite $\tau + c$ but cannot separate its two components.

We make the ambiguity explicit.
Fix any function $g : \mathcal{X} \to \mathbb{R}$ and set $\tilde\tau(x) := \tau(x) + g(x)$ and $\tilde c(x) := c(x) - g(x)$.
The sum is preserved: $\tilde\tau(x) + \tilde c(x) = \tau(x) + c(x)$ for every $x$.
Substitution of $(\tilde\tau, \tilde c)$ for $(\tau, c)$ into \eqref{eq:rwd-treated} reproduces the same RWD conditional means.
The two parameter pairs are observationally equivalent on the RWD slice.

Two natural choices of $g$ illustrate the ambiguity.
Take $g \equiv -\tau$.
Then $\tilde\tau \equiv 0$ and $\tilde c \equiv \tau + c$: the entire RWD contrast is attributed to confounding bias, with no causal effect.
Take $g \equiv c$.
Then $\tilde\tau \equiv \tau + c$ and $\tilde c \equiv 0$: the entire RWD contrast is attributed to a causal effect, under the false premise that the RWD is unconfounded.
The two interpretations are indistinguishable from the RWD data alone.
The non-identification is structural rather than statistical.
Larger RWD samples cannot resolve it.
External information that identifies one of the two functions is required.
\qed

\newpage
\section{The hierarchical $m_0$ prior as a meta-analytic-predictive prior}\label{sm:map-m0}

The meta-analytic-predictive (MAP) prior of \citet{neuenschwander2010} is a hierarchical Bayesian construction for borrowing strength between studies.
We argue that the main-text specification \eqref{eq:m0-decomp-prop} of $m_0(X, S)$ is a function-valued, heterogeneous MAP prior.
The symmetric form of this prior is the canonical MAP construction.
The main text uses its two-study instance, anchored at the RCT, which we derive below.

The MAP prior assumes a Gaussian hierarchical structure across $K$ studies with study-specific parameters $\theta_1, \ldots, \theta_K$:
\begin{align}
\theta_k \mid \theta^*, \sigma_d^2 &\;\sim\; N(\theta^*, \sigma_d^2), \qquad k = 1, \ldots, K, \\
\theta^* &\;\sim\; p(\theta^*), \\
\sigma_d &\;\sim\; p(\sigma_d).
\label{eq:map-scalar}
\end{align}
The shared mean $\theta^*$ is the borrowing target.
The heterogeneity variance $\sigma_d^2$ controls how strongly each $\theta_k$ is shrunk towards $\theta^*$.
Small $\sigma_d$ enforces near-equality of the $\theta_k$, i.e.\ strong borrowing.
Large $\sigma_d$ allows the $\theta_k$ to drift apart, i.e.\ weak borrowing.
The hierarchical structure lets the data determine the degree of borrowing through the posterior of $\sigma_d$.

We restrict the MAP prior to two studies before extending it to functions.
Write $\theta_s = \theta^* + d_s$ with $d_s \sim \mathcal{N}(0, \sigma_d^2)$ for the RWD ($s = 0$) and the RCT ($s = 1$).
The model carries three latent quantities, $\theta^*$, $d_0$, and $d_1$, but the two studies determine only the two parameters $\theta_0$ and $\theta_1$.
The overall level of $\theta^*$ is therefore not separately identified from the study-specific deviations.
We resolve this by anchoring at the RCT.
Take a vague prior on the shared mean, $\theta^* \sim \mathcal{N}(0, \kappa^2)$ with $\kappa^2 \to \infty$.
The pair $(\theta_0, \theta_1)$ is then jointly Gaussian with $\mathrm{Var}(\theta_s) = \kappa^2 + \sigma_d^2$ and $\mathrm{Cov}(\theta_0, \theta_1) = \kappa^2$.
The contrast satisfies $\theta_0 - \theta_1 = d_0 - d_1 \sim \mathcal{N}(0,\, 2\sigma_d^2)$.
We condition on the RCT parameter $\theta_1$ and let $\kappa^2 \to \infty$.
The regression coefficient then tends to one and the conditional variance tends to $2\sigma_d^2$:
\begin{equation}
\theta_0 \mid \theta_1 \;\sim\; N\!\left(\theta_1,\; 2\sigma_d^2\right).
\label{eq:map-conditional}
\end{equation}
The shared level $\theta^*$ has dropped out, and the RCT parameter $\theta_1$ now serves as the borrowing target.
This anchoring at the RCT is a reduction forced by having two studies and a non-informative level, not an extra assumption.
Equation \eqref{eq:map-conditional} also reads as an informative prior that centres the RWD parameter on the RCT parameter, with a heterogeneity scale that controls the borrowing \citep{schmidli2014robust}.

We now extend the scalar reduction to functions.
The study-specific parameter becomes the prognostic function evaluated at source $s$, $\theta_s(X) := m_0(X, s)$, and the Gaussian deviation becomes a mean-zero BART deviation.
The conditional form \eqref{eq:map-conditional} then gives the main-text specification \eqref{eq:m0-decomp-prop}:
\begin{equation}
m_0(X, S) \;=\; m_0^{\mathrm{sh}}(X) + (1 - S) \cdot d(X).
\end{equation}
We set the shared function equal to the RCT prognosis, $m_0^{\mathrm{sh}} := m_0(\cdot, 1)$, and write the single RWD deviation as $d$, with $d \sim \mathrm{BART}(\sigma_{h, d})$ and mean zero.
Here $\mathrm{BART}(\sigma_h)$ denotes a BART prior indexed by its leaf scale $\sigma_h$, suppressing the tree-structure arguments of the full notation $\mathrm{BART}(J, k, \alpha, \beta)$.
The deviation $d$ carries the between-source heterogeneity, and the prior is heterogeneous because $d$ varies with $X$.
By \eqref{eq:map-conditional} the scale of $d$ is the symmetric deviation scale inflated by $\sqrt{2}$, which we absorb into the calibration of $k_d$ below.
In the functional model we do not take the vague-mean limit literally.
We instead assign $m_0^{\mathrm{sh}}$ its own $\mathrm{BART}(200, \ldots)$ prior as a regularised borrowing target, and use the limit only to justify the anchoring.
For more than two sources, or when neither source is a natural reference, we keep a separate mean-zero deviation $d_s$ for each source, which retains the same identifiability of $\tau$ and $c$.

We treat the borrowing scale as a calibrated hyperparameter rather than sampling it.
The leaf scale of the deviation forest is set by the Chipman--George--McCulloch rule \citep{chipman2010bart}:
\begin{equation}
\sigma_{h, d} \;=\; \frac{k_d}{2\sqrt{J_d}},
\label{eq:map-sigma-d}
\end{equation}
where $J_d$ is the number of trees in the deviation forest and $k_d > 0$ is the tuning constant of the Chipman calibration.
The marginal standard deviation of $d(X)$ is then $\sqrt{J_d}\, \sigma_{h, d} = k_d / 2$, independent of $J_d$.
So $k_d$ is the borrowing-strength dial.
Small $k_d$ shrinks $d$ towards zero and induces strong borrowing of the RWD prognostic towards the RCT prognostic.
Large $k_d$ relaxes the prior and lets $d$ depart further from zero.
We take $k_d = 1$ by default, in line with the uniform Chipman calibration of the main text, and the $\sqrt{2}$ factor from the anchored reduction \eqref{eq:map-conditional} is absorbed into this choice.
A fully Bayesian alternative is to put a hyperprior on the borrowing scale as in the standard and robust MAP \citep{neuenschwander2010, schmidli2014robust}; this trades the simplicity of the BART convention for additional data-driven adaptation of the borrowing strength.

The standard MAP \eqref{eq:map-scalar} assumes Gaussian deviations, which is restrictive when one or more studies are outliers.
\citet{hupf2021bayesian} extend MAP to flexible deviations through a Dirichlet-process-based prior on the heterogeneity distribution.
The BART deviation in our prior is flexible in a similar spirit, though along a different axis.
\citet{hupf2021bayesian} make the distribution of deviations across studies flexible, which guards against an outlying study.
Our deviation instead lets the shape of $d$ vary flexibly with the covariates $X$, while retaining the borrowing target $m_0^{\mathrm{sh}}$.

The construction justifies the two-forest decomposition of the baseline prognosis.
A single BART on $m_0(X, S)$ with $S$ as a covariate cannot encode the preference for borrowing, and two independent baselines would imply a needlessly diffuse prior on their difference.
The shared-plus-deviation form instead places the prior directly on the between-source difference and tunes the borrowing through the deviation scale.

\newpage
\section{Posterior computation}\label{sm:posterior-computation}

We sample the posterior by a blocked Gibbs sampler.
We recap the model, summarise one iteration of the outer Gibbs sampler, and derive the update for each parameter block in turn.
We close with the calibration of the hyperparameters specific to the error model.

\subsection{Model recap and augmented representation}

The Bayesian fusion forest of the main text decomposes the log survival time as:
\begin{equation}
\log T_i \;=\; m_0^{\mathrm{sh}}(x_i) + (1 - s_i)\, d(x_i) + a_i\, \tau(x_i) + (1 - s_i)\, a_i\, c(x_i) + \varepsilon_i,
\label{eq:sm-aft}
\end{equation}
with $s_i \in \{0, 1\}$ the source indicator ($s_i = 1$ for RCT, $s_i = 0$ for RWD), $a_i \in \{0, 1\}$ the treatment indicator, and $x_i \in \mathbb{R}^p$ the covariate vector.
The four regression functions $f \in \{\mathrm{sh}, d, \tau, c\}$ carry independent BART priors.
The errors satisfy $\varepsilon_i \mid S_i = s \sim F_s$, with $F_s$ a centred location mixture of Gaussians:
\begin{equation}
\varepsilon_i \mid S_i = s,\, G_s,\, \sigma_s \;\sim\; \int \frac{1}{\sigma_s}\, \phi\!\left(\frac{w - \theta}{\sigma_s}\right) dG_s(\theta), \qquad G_s = \sum_{k} \pi_{sk}\, \delta_{\theta_k^* - \mu_s},
\end{equation}
with shared atoms $\theta_k^*$, source-specific weights $\bm{\pi}_s = (\pi_{sk})$, and source-specific shifts $\mu_s = \sum_k \pi_{sk} \theta_k^*$ that enforce $\E[\varepsilon_i \mid S_i = s] = 0$.
The mixing measures are tied across sources through a hierarchical Dirichlet process with concentrations $\gamma$ (top-level) and $M_0, M_1$ (source-specific), and per-source residual scales $\sigma_s$.

We observe each subject as a triple $(L_i, R_i, \delta_i)$ with $T_i \in [L_i, R_i]$ and $\delta_i \in \{0, 1, 2\}$ encoding the censoring type.
The value $\delta_i = 1$ corresponds to exact observation ($L_i = R_i = T_i$).
The value $\delta_i = 0$ corresponds to right censoring ($L_i = C_i$, $R_i = \infty$).
The value $\delta_i = 2$ corresponds to interval censoring ($0 \leq L_i < R_i < \infty$).
The sampler treats $\log T_i$ as a latent quantity for every subject with $\delta_i \neq 1$, and augments it at the end of each iteration.

For each observation we define the partial residual on the augmented log-time scale:
\begin{equation}
R_i \;:=\; \log T_i - m_0^{\mathrm{sh}}(x_i) - (1 - s_i)\, d(x_i) - a_i\, \tau(x_i) - a_i(1 - s_i)\, c(x_i).
\label{eq:sm-Ri}
\end{equation}
The residual collects the contribution of the error $\varepsilon_i$ at the current draw of the regression functions.

For computation we truncate the top-level stick-breaking at a finite level $K$ \citep{sethuraman1994constructive, ishwaran2001gibbs}.
The top-level weights $\bm{\beta} = (\beta_1, \ldots, \beta_K)$ are constructed by $\beta_k = u_k \prod_{l < k}(1 - u_l)$ with $u_k \stackrel{\mathrm{iid}}{\sim} \mathrm{Beta}(1, \gamma)$.
We set $u_K = 1$ to truncate at level $K$, so that $\beta_K = 1 - \sum_{k < K} \beta_k$ and $\beta_k = 0$ for $k > K$.
The source-specific weights then have a finite-dimensional Dirichlet prior on the $K$-simplex, and the approximation becomes exact as $K \to \infty$.
We introduce latent cluster assignments $Z_i \in \{1, \ldots, K\}$ such that the latent location of observation $i$ is $\theta_{Z_i}^* - \mu_{s_i}$.
The full hierarchical model after truncation reads:
\begin{align}
\log T_i \mid \cdot &\;\sim\; N\!\left(\eta_i,\, \sigma_{s_i}^2\right),\\
\eta_i &\;=\; m_0^{\mathrm{sh}}(x_i) + (1 - s_i)\, d(x_i) + a_i\, \tau(x_i) + a_i(1 - s_i)\, c(x_i) + \theta_{Z_i}^* - \mu_{s_i},
  \label{eq:sm-eta}\\
m_0^{\mathrm{sh}} &\;\sim\; \mathrm{BART}(J_{\mathrm{sh}}, k_{\mathrm{sh}}, \alpha, \beta),\\
d &\;\sim\; \mathrm{BART}(J_d, k_d, \alpha, \beta),\\
\tau &\;\sim\; \mathrm{BART}(J_\tau, k_\tau, \alpha_\tau, \beta_\tau),\\
c &\;\sim\; \mathrm{BART}(J_c, k_c, \alpha_c, \beta_c),\\
Z_i \mid \bm{\pi}_{s_i} &\;\sim\; \mathrm{Categorical}(\bm{\pi}_{s_i}),\\
\bm{\pi}_s \mid \bm{\beta},\, M_s &\;\sim\; \mathrm{Dirichlet}(M_s \beta_1, \ldots, M_s \beta_K), \qquad s \in \{0, 1\},\\
\bm{\beta} \mid \gamma &\;\sim\; \mathrm{Stick}_K(\gamma),\\
\theta_k^* &\;\stackrel{\mathrm{iid}}{\sim}\; \mathcal{N}(0, \sigma_\theta^2), \qquad k = 1, \ldots, K,\\
\sigma_s^2 &\;\sim\; \nu \lambda / \chi^2_\nu, \qquad s \in \{0, 1\},\\
\gamma &\;\sim\; \mathrm{Gamma}(a_\gamma, b_\gamma),\\
M_s &\;\sim\; \mathrm{Gamma}(a_M, b_M), \qquad s \in \{0, 1\},
\end{align}
where $\mathrm{Stick}_K(\gamma)$ is the truncated stick-breaking construction given above.
We use a default $K = 50$ and track the largest occupied component index across iterations to confirm that it stays well below $K$.
In our simulation studies the number of occupied components did not exceed 25, so the default $K = 50$ leaves substantial headroom.

\subsection{Outer Gibbs sampler}

One iteration of the sampler consists of three blocks executed in order: updates of the four forests $\{\mathrm{sh}, d, \tau, c\}$, an update of the HDPM error block, and augmentation of the censored event times.
We update the forests sequentially against partial residuals.
We update the HDPM error block as a single unit because its conditional structure is internally coupled.
The augmentation step at the end of the iteration updates the latent $\log T_i$ for use in the next iteration.
Algorithm~\ref{alg:sm-outer} summarises one iteration.

\begin{algorithm}[H]
\caption{Outer Gibbs sampler, one iteration.}
\label{alg:sm-outer}
\KwIn{Current forests $(\mathcal{T}_j^f, \mathcal{H}_j^f)$ for $f \in \{\mathrm{sh}, d, \tau, c\}$; current augmented event times $\log T_i$; HDPM state $(\bm{\theta}^*, \bm{\beta}, \bm{\pi}_0, \bm{\pi}_1, \gamma, M_0, M_1, \sigma_0^2, \sigma_1^2)$.}
\vspace{1ex}

\ForEach{forest $f \in \{\mathrm{sh}, d, \tau, c\}$}{
  \For{$j = 1$ \KwTo $J_f$}{
    Update $(\mathcal{T}_j^f, \mathcal{H}_j^f)$ by one BART backfitting step on the partial residual $R_i^f$, $i \in \mathcal{I}_f$\;
  }
}
\vspace{1ex}

Update the HDPM error block (Algorithm~\ref{alg:sm-hdp})\;
\vspace{1ex}

\For{$i = 1$ \KwTo $n$ with $\delta_i \neq 1$}{
  Draw $\log T_i$ from its truncated normal full conditional\;
}
\end{algorithm}

\subsection{Forest updates}

We update the four forests sequentially by extended Bayesian backfitting \citep{hastie2000bayesian, chipman2010bart}.
We define a partial residual $R_i^f$ on the active subset $\mathcal{I}_f$ of observations for each forest $f \in \{\mathrm{sh}, d, \tau, c\}$:
\begin{align}
R_i^{m_0^{\mathrm{sh}}} &= \log T_i - (1-s_i)\, d(x_i) - a_i\, \tau(x_i) - a_i(1-s_i)\, c(x_i) - (\theta_{Z_i}^* - \mu_{s_i}),\\
R_i^{d} &= \log T_i - m_0^{\mathrm{sh}}(x_i) - a_i\, \tau(x_i) - a_i\, c(x_i) - (\theta_{Z_i}^* - \mu_0),\\
R_i^{\tau} &= \log T_i - m_0^{\mathrm{sh}}(x_i) - (1-s_i)\, d(x_i) - (1-s_i)\, c(x_i) - (\theta_{Z_i}^* - \mu_{s_i}),\\
R_i^{c} &= \log T_i - m_0^{\mathrm{sh}}(x_i) - d(x_i) - \tau(x_i) - (\theta_{Z_i}^* - \mu_0),
\end{align}
with active subsets
\begin{equation}
\mathcal{I}_{m_0^{\mathrm{sh}}} = \{1, \ldots, n\},\quad
\mathcal{I}_d = \{i : s_i = 0\},\quad
\mathcal{I}_\tau = \{i : a_i = 1\},\quad
\mathcal{I}_c = \{i : a_i = 1,\, s_i = 0\}.
\end{equation}
Each $R_i^f$ subtracts the contributions of the other three forests and the current cluster shift $\theta_{Z_i}^* - \mu_{s_i}$ from $\log T_i$, restricted to the subset on which $f$ is active.
Conditional on the cluster assignment $Z_i$ and the atoms $\theta^*$, the residual $R_i^f$ is Gaussian with mean $f(x_i)$ and variance $\sigma_{s_i}^2$ for $i \in \mathcal{I}_f$. The BART step on forest $f$ operates on this Gaussian working response.

Within forest $f$, we update each tree by a standard BART backfitting step on $R_i^f$.
We update the tree structure $\mathcal{T}_j^f$ by a Metropolis-Hastings step with birth-death proposals \citep{chipman1998bayesian, kapelner2016bartmachine}.
We then draw the leaf step heights $\mathcal{H}_j^f$ from their conjugate Gaussian posterior.
The Gaussian leaf prior is conjugate to the Gaussian working response, and the posterior mean and variance at each terminal node are available in closed form.

\subsection{HDPM error block}

The HDPM error block updates the cluster assignments, atoms, top-level sticks, source-specific weights, concentration parameters, and per-source residual variances.
All full conditionals are conjugate.
Throughout, we write $\mid \cdot$ for conditioning on the data and on the current values of all other parameters.
Algorithm~\ref{alg:sm-hdp} summarises one update of the block.

We sample the cluster assignments from
\begin{equation}
\p(Z_i = k \mid \cdot) \;\propto\; \pi_{s_i, k}\, \phi\!\left\{\frac{R_i - (\theta_k^* - \mu_{s_i})}{\sigma_{s_i}}\right\}, \qquad k = 1, \ldots, K.
\label{eq:sm-cluster}
\end{equation}

We sample the atoms in the unconstrained parameterisation and reapply the centring deterministically \citep{yang2010semiparametric}.
Let $n_{sk} = \sum_i \1\{s_i = s, Z_i = k\}$ and $\tilde R_{sk} = \sum_{i : s_i = s,\, Z_i = k} (R_i + \mu_s)$.
The two sources pool through the precision weights $\sigma_s^{-2}$ in both the mean and the variance of the conjugate update:
\begin{equation}
\theta_k^* \mid \cdot \;\sim\; N\!\left(\frac{\sum_{s} \sigma_s^{-2}\, \tilde R_{sk}}{\sigma_\theta^{-2} + \sum_{s} \sigma_s^{-2}\, n_{sk}},\; \bigl(\sigma_\theta^{-2} + \sum_{s} \sigma_s^{-2}\, n_{sk}\bigr)^{-1}\right), \qquad k = 1, \ldots, K.
\label{eq:sm-atoms}
\end{equation}
We then recompute $\mu_s = \sum_k \pi_{sk} \theta_k^*$ and the centred atoms $\theta_k^* - \mu_s$.

We update the top-level stick weights by the auxiliary table-count augmentation of \citet{teh2006hierarchical}.
For each $(s, k)$ with $n_{sk} > 0$ we sample the auxiliary table count via the Antoniak representation \citep{antoniak1974mixtures}:
\begin{equation}
t_{sk} \;=\; \sum_{j=1}^{n_{sk}} B_{sk,j}, \qquad B_{sk,j} \;\sim\; \mathrm{Bernoulli}\!\left(\frac{M_s \beta_k}{M_s \beta_k + j - 1}\right),
\end{equation}
and we set $t_{sk} = 0$ when $n_{sk} = 0$.
With $t_{\cdot k} = t_{0k} + t_{1k}$, we update the truncated sticks by
\begin{equation}
u_k \mid \cdot \;\sim\; \mathrm{Beta}\!\left(1 + t_{\cdot k},\; \gamma + \sum_{l > k} t_{\cdot l}\right), \qquad k = 1, \ldots, K - 1,
\end{equation}
with $u_K = 1$ and $\beta_k = u_k \prod_{l < k}(1 - u_l)$.
The source-specific weights then follow
\begin{equation}
\bm{\pi}_s \mid \cdot \;\sim\; \mathrm{Dirichlet}(M_s \beta_1 + n_{s1},\, \ldots,\, M_s \beta_K + n_{sK}), \qquad s \in \{0, 1\},
\end{equation}
after which we recompute $\mu_s$ and the centred atoms.

We update the concentration parameters $\gamma$ and $M_0, M_1$ by the auxiliary-variable scheme of \citet{escobar1995bayesian}. See \citet[Appendix A]{teh2006hierarchical} for the derivation.
Let $K^*$ denote the number of occupied top-level components and $t_{\cdot\cdot} = \sum_{s, k} t_{sk}$ the total table count.
We draw $\eta_\gamma \mid \gamma, t_{\cdot\cdot} \sim \mathrm{Beta}(\gamma + 1, t_{\cdot\cdot})$ and then sample $\gamma$ from a mixture of two Gamma distributions with mixing weight $\omega_\gamma$:
\begin{equation}
\gamma \mid \cdot \;\sim\; \omega_\gamma\, \mathrm{Gamma}(a_\gamma + K^*,\, b_\gamma - \log \eta_\gamma) + (1 - \omega_\gamma)\, \mathrm{Gamma}(a_\gamma + K^* - 1,\, b_\gamma - \log \eta_\gamma),
\end{equation}
with $\omega_\gamma / (1 - \omega_\gamma) = (a_\gamma + K^* - 1) / [t_{\cdot\cdot}(b_\gamma - \log \eta_\gamma)]$.
The analogous update for $M_s$ replaces $K^*$ by the source-specific table count $t_{s\cdot} = \sum_k t_{sk}$ and uses $\eta_{M, s} \mid M_s, n_s \sim \mathrm{Beta}(M_s + 1, n_s)$.

The conjugate update of the per-source residual variance closes the block:
\begin{equation}
\sigma_s^2 \mid \cdot \;\sim\; \mathrm{InverseGamma}\!\left(\frac{\nu + n_s}{2},\; \frac{\nu \lambda + \mathrm{SS}_s}{2}\right), \qquad \mathrm{SS}_s = \sum_{i : S_i = s} \bigl\{R_i - (\theta_{Z_i}^* - \mu_s)\bigr\}^2, \qquad s \in \{0, 1\}.
\label{eq:sm-sigma}
\end{equation}

\begin{algorithm}[H]
\caption{Update of the HDPM error block.}
\label{alg:sm-hdp}
\KwIn{Partial residuals $R_i$ from \eqref{eq:sm-Ri}; current atoms $\bm{\theta}^*$, sticks $\bm{\beta}$, source weights $\bm{\pi}_0, \bm{\pi}_1$, concentrations $\gamma, M_0, M_1$, scales $\sigma_0^2, \sigma_1^2$.}
\vspace{1ex}

\For{$i = 1$ \KwTo $n$}{
  Sample cluster assignment $Z_i$ from the categorical full conditional \eqref{eq:sm-cluster}\;
}
\vspace{1ex}

\For{$k = 1$ \KwTo $K$}{
  Update the unconstrained atom $\theta_k^*$ from its conjugate Gaussian posterior \eqref{eq:sm-atoms}\;
}
Recompute $\mu_s = \sum_k \pi_{sk} \theta_k^*$ and the centred atoms\;
\vspace{1ex}

Sample auxiliary table counts $t_{sk}$ via the Antoniak representation\;
Update top-level sticks $u_k$ from Beta posteriors\;
Recompute $\beta_k = u_k \prod_{l < k}(1 - u_l)$\;
\vspace{1ex}

\ForEach{$s \in \{0, 1\}$}{
  Update source weights $\bm{\pi}_s$ from a Dirichlet posterior\;
}
Recompute $\mu_s$ and the centred atoms\;
\vspace{1ex}

Update concentrations $\gamma, M_0, M_1$ via the Escobar-West auxiliary-variable scheme\;
\vspace{1ex}

\ForEach{$s \in \{0, 1\}$}{
  Update the residual scale $\sigma_s^2$ from its inverse-gamma posterior \eqref{eq:sm-sigma}\;
}
\end{algorithm}

\subsection{Data augmentation for censored event times}

Each iteration closes with data augmentation of the unobserved event times for the censored observations \citep{tanner1987}.
Subjects with $\delta_i = 1$ contribute the observed $\log T_i = \log L_i = \log R_i$ and need no augmentation.
Subjects with $\delta_i = 0$ have a right-censored event time and we draw
\begin{equation}
\log T_i \mid \cdot \;\sim\; \mathrm{TruncatedNormal}\!\left(\eta_i,\, \sigma_{s_i}^2;\, [\log L_i, \infty)\right),
\end{equation}
where $\eta_i$ is the conditional mean of $\log T_i$ defined in \eqref{eq:sm-eta}.
Subjects with $\delta_i = 2$ have an interval-censored event time and we draw
\begin{equation}
\log T_i \mid \cdot \;\sim\; \mathrm{TruncatedNormal}\!\left(\eta_i,\, \sigma_{s_i}^2;\, [\log L_i,\, \log R_i]\right).
\end{equation}
The augmented values serve as the working response for the next iteration.

\subsection{Hyperparameters of the nonparametric error distribution}

The nonparametric error model carries three sets of hyperparameters: the per-source residual scales $\sigma_s^2$, the concentration parameters $\gamma, M_0, M_1$, and the base-measure variance $\sigma_\theta^2$.
We collect their priors and calibration here.

We place a scaled-inverse-$\chi^2$ prior $\sigma_s^2 \sim \nu \lambda / \chi^2_\nu$ on the per-source residual scales, with $\nu = 3$ and $\lambda$ calibrated to the empirical residual variance \citep{chipman2010bart}.
We update each $\sigma_s^2$ by the inverse-gamma conjugate step \eqref{eq:sm-sigma}.

We treat the concentration parameters $\gamma, M_0, M_1$ as unknown and estimate them.
We place independent Gamma hyperpriors $\gamma \sim \mathrm{Gamma}(a_\gamma, b_\gamma)$ and $M_s \sim \mathrm{Gamma}(a_M, b_M)$, with $a_\gamma = a_M = 2$ and $b_\gamma = b_M = 0.1$.
This places prior mean $20$ and prior mode $10$ on each concentration.
We update them by the Escobar-West auxiliary-variable scheme of the HDPM error block.

The base-measure variance $\sigma_\theta^2$ of the unconstrained atoms is calibrated by a Yamato-type approximation \citep{yamato1984characteristic}.
The quantity $\sum_k \pi_{sk}(\theta_k^* - \mu_s)^2 / \sigma_\theta^2$ is approximately $\chi^2_1$.
We match the induced prior on the marginal residual variance $\mathrm{Var}(\varepsilon \mid S = s)$ in tail probability to a preliminary parametric estimate $\hat\sigma_\varepsilon^2$.
We apply the matching within each source separately, with a default tail probability $q = 0.5$.

\subsection{BART hyperparameters and initialisation}

The BART tree-prior hyperparameters $(J_f, k_f, \alpha_f, \beta_f)$ are set as in the main text.
We initialise the BART forests by the standard \citet{chipman2010bart} initialisation.
We initialise the cluster labels by $k$-means on the residuals of a preliminary parametric log-normal AFT fit.
We initialise the HDPM weights at $\bm{\beta} = (1/K, \ldots, 1/K)$, $\bm{\pi}_s = \bm{\beta}$, and the concentrations at $\gamma = M_0 = M_1 = 1$.

\newpage
\section{Additional results for the simulation study}\label{sm:sim-additional}

\begin{figure}[!b]
  \centering
  \includegraphics[width=0.8\textwidth]{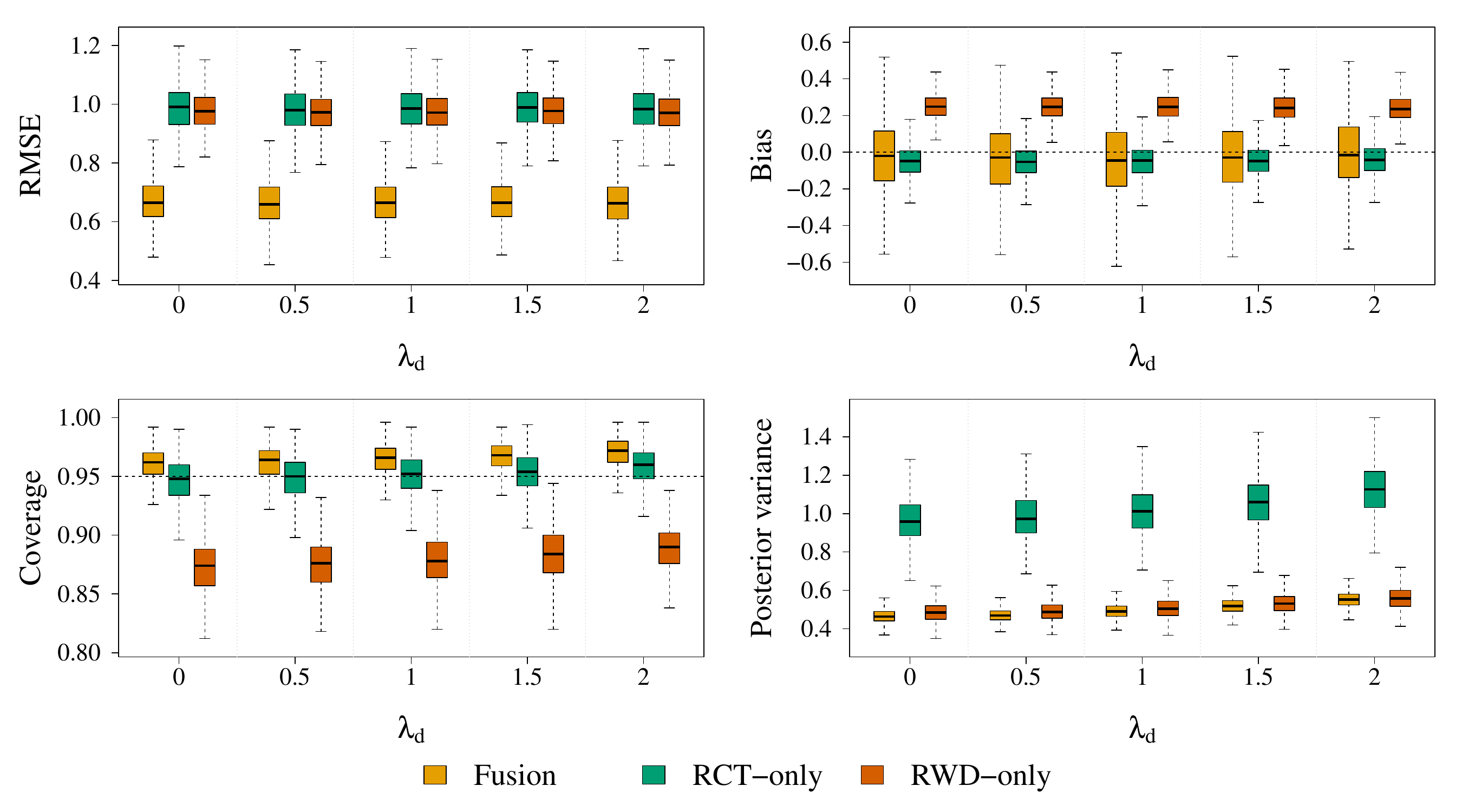}
  \caption{CATE metrics under varying between-source heterogeneity $\lambda_d$, with the unmeasured confounding held at $\lambda_u = 1$. Panels, methods, and reference lines as in Figure~\ref{fig:sim-lambda-u} of the main text.}
  \label{fig:sim-lambda-d}
\end{figure}

We report the simulation under varying between-source heterogeneity $\lambda_d$ here.
Figure~\ref{fig:sim-lambda-d} gives the CATE metrics across the grid of $\lambda_d$.
The main text discusses these results.

We give the full comparison with the deep-learning competitors here.
Table~\ref{tab:sim-deepaft} reports each metric across the $(\lambda_d, \lambda_u)$ grid, with $\lambda_d, \lambda_u \in \{0, 1, 2\}$.
The grid restricts to the deep network because it is the strongest of the four alternatives (Table~\ref{tab:sim-competitors} of the main text).
The Bayesian fusion forest attains the lowest RMSE at every grid point.

\begin{sidewaystable}
  \centering
  \caption{CATE metrics for the Bayesian fusion forest (BFF) and the four
    deepAFT competitors across the $(\lambda_d, \lambda_u)$ grid. Each entry
    is the mean over simulation replicates, with the 2.5--97.5\% range in
    brackets. R and P denote the RCT-only and naive-pool fits; S and T denote
    the S-learner and T-learner.}
  \label{tab:sim-deepaft}
  \scriptsize
  \setlength{\tabcolsep}{4pt}
  \begin{tabular}{cc ccccc}
    \toprule
    $\lambda_d$ & $\lambda_u$ & BFF & DNN (RCT, S) & DNN (RCT, T) & DNN (pool, S) & DNN (pool, T) \\
    \midrule
    \multicolumn{7}{l}{\emph{RMSE}}\\
    0 & 0 & 0.68 [0.53;0.87] & 1.11 [0.98;1.28] & 1.24 [0.89;1.75] & 1.08 [0.94;1.27] & 1.11 [0.70;3.04] \\
    1 & 0 & 0.68 [0.53;0.85] & 1.11 [0.98;1.27] & 1.25 [0.88;1.74] & 1.08 [0.93;1.27] & 1.11 [0.73;2.94] \\
    2 & 0 & 0.67 [0.52;0.85] & 1.12 [0.98;1.29] & 1.27 [0.89;1.84] & 1.09 [0.95;1.27] & 1.22 [0.83;3.06] \\
    0 & 1 & 0.68 [0.54;0.85] & 1.11 [0.97;1.29] & 1.23 [0.91;1.74] & 1.10 [0.96;1.27] & 1.09 [0.71;3.12] \\
    1 & 1 & 0.68 [0.52;0.86] & 1.12 [0.98;1.26] & 1.25 [0.89;1.77] & 1.10 [0.96;1.28] & 1.13 [0.74;3.02] \\
    2 & 1 & 0.68 [0.53;0.84] & 1.13 [0.98;1.28] & 1.26 [0.88;1.86] & 1.12 [0.96;1.28] & 1.23 [0.83;3.13] \\
    0 & 2 & 0.68 [0.53;0.86] & 1.11 [0.97;1.26] & 1.25 [0.91;1.72] & 1.16 [1.02;1.33] & 1.12 [0.72;3.16] \\
    1 & 2 & 0.69 [0.54;0.87] & 1.12 [0.99;1.27] & 1.25 [0.90;1.76] & 1.17 [1.01;1.33] & 1.19 [0.78;3.26] \\
    2 & 2 & 0.68 [0.53;0.87] & 1.12 [0.98;1.29] & 1.24 [0.87;1.85] & 1.17 [1.01;1.33] & 1.25 [0.82;3.20] \\
    \midrule
    \multicolumn{7}{l}{\emph{Bias}}\\
    0 & 0 & $-0.01$ [$-0.42$;0.38] & $-0.17$ [$-0.55$;0.23] & $-0.21$ [$-0.70$;0.33] & $-0.13$ [$-0.39$;0.14] & $-0.02$ [$-0.47$;0.74] \\
    1 & 0 & $-0.03$ [$-0.44$;0.36] & $-0.20$ [$-0.53$;0.19] & $-0.23$ [$-0.70$;0.27] & $-0.13$ [$-0.42$;0.14] & 0.01 [$-0.48$;0.76] \\
    2 & 0 & $-0.04$ [$-0.46$;0.36] & $-0.20$ [$-0.53$;0.20] & $-0.28$ [$-0.81$;0.28] & $-0.12$ [$-0.40$;0.18] & $-0.04$ [$-0.58$;0.88] \\
    0 & 1 & $-0.03$ [$-0.42$;0.37] & $-0.18$ [$-0.52$;0.20] & $-0.23$ [$-0.72$;0.29] & 0.11 [$-0.14$;0.39] & 0.09 [$-0.36$;0.92] \\
    1 & 1 & $-0.03$ [$-0.45$;0.37] & $-0.19$ [$-0.54$;0.21] & $-0.25$ [$-0.77$;0.26] & 0.11 [$-0.16$;0.38] & 0.10 [$-0.39$;0.91] \\
    2 & 1 & $-0.01$ [$-0.42$;0.41] & $-0.20$ [$-0.54$;0.18] & $-0.25$ [$-0.80$;0.32] & 0.10 [$-0.17$;0.38] & 0.06 [$-0.51$;0.99] \\
    0 & 2 & $-0.03$ [$-0.43$;0.37] & $-0.20$ [$-0.55$;0.16] & $-0.25$ [$-0.75$;0.26] & 0.36 [$-0.03$;0.65] & 0.19 [$-0.30$;1.08] \\
    1 & 2 & $-0.02$ [$-0.43$;0.39] & $-0.19$ [$-0.53$;0.21] & $-0.24$ [$-0.76$;0.31] & 0.35 [$-0.01$;0.65] & 0.19 [$-0.35$;1.25] \\
    2 & 2 & 0.00 [$-0.43$;0.42] & $-0.19$ [$-0.53$;0.23] & $-0.24$ [$-0.77$;0.33] & 0.32 [$-0.05$;0.65] & 0.21 [$-0.43$;1.25] \\
    \midrule
    \multicolumn{7}{l}{\emph{Coverage}}\\
    0 & 0 & 0.96 [0.92;0.98] & 0.38 [0.28;0.50] & 0.91 [0.76;0.99] & 0.41 [0.31;0.52] & 0.99 [0.97;1.00] \\
    1 & 0 & 0.96 [0.93;0.98] & 0.37 [0.27;0.48] & 0.91 [0.78;0.99] & 0.42 [0.32;0.54] & 0.99 [0.97;1.00] \\
    2 & 0 & 0.97 [0.94;0.99] & 0.37 [0.27;0.48] & 0.92 [0.79;0.99] & 0.43 [0.34;0.54] & 0.99 [0.97;1.00] \\
    0 & 1 & 0.96 [0.93;0.98] & 0.38 [0.27;0.48] & 0.91 [0.77;0.99] & 0.44 [0.33;0.56] & 0.99 [0.96;1.00] \\
    1 & 1 & 0.96 [0.93;0.99] & 0.38 [0.28;0.49] & 0.91 [0.78;0.99] & 0.44 [0.34;0.56] & 0.99 [0.97;1.00] \\
    2 & 1 & 0.97 [0.94;0.99] & 0.37 [0.27;0.48] & 0.92 [0.79;0.99] & 0.45 [0.34;0.56] & 0.99 [0.96;1.00] \\
    0 & 2 & 0.96 [0.93;0.99] & 0.37 [0.27;0.49] & 0.91 [0.77;0.99] & 0.45 [0.36;0.56] & 0.99 [0.96;1.00] \\
    1 & 2 & 0.96 [0.93;0.99] & 0.37 [0.27;0.48] & 0.91 [0.79;0.99] & 0.45 [0.35;0.58] & 0.99 [0.96;1.00] \\
    2 & 2 & 0.97 [0.94;0.99] & 0.37 [0.27;0.49] & 0.93 [0.78;0.99] & 0.45 [0.35;0.57] & 0.99 [0.96;1.00] \\
    \midrule
    \multicolumn{7}{l}{\emph{Posterior variance}}\\
    0 & 0 & 0.48 [0.41;0.56] & 0.07 [0.04;0.12] & 1.15 [0.74;1.94] & 0.12 [0.05;0.38] & 3.96 [0.97;11.76] \\
    1 & 0 & 0.50 [0.44;0.58] & 0.07 [0.04;0.14] & 1.18 [0.74;2.06] & 0.12 [0.05;0.34] & 4.19 [1.05;11.48] \\
    2 & 0 & 0.56 [0.49;0.65] & 0.07 [0.04;0.15] & 1.35 [0.74;2.98] & 0.13 [0.05;0.42] & 4.72 [1.18;13.42] \\
    0 & 1 & 0.49 [0.42;0.57] & 0.07 [0.04;0.11] & 1.10 [0.71;1.79] & 0.12 [0.05;0.40] & 3.72 [1.02;10.24] \\
    1 & 1 & 0.52 [0.44;0.60] & 0.07 [0.04;0.11] & 1.13 [0.74;1.93] & 0.13 [0.05;0.42] & 4.13 [1.09;11.12] \\
    2 & 1 & 0.58 [0.50;0.67] & 0.07 [0.04;0.12] & 1.21 [0.76;2.35] & 0.13 [0.05;0.45] & 5.19 [1.24;14.93] \\
    0 & 2 & 0.52 [0.45;0.61] & 0.07 [0.04;0.10] & 1.08 [0.72;1.65] & 0.15 [0.06;0.45] & 3.60 [1.08;10.63] \\
    1 & 2 & 0.55 [0.46;0.63] & 0.06 [0.04;0.10] & 1.09 [0.74;1.62] & 0.15 [0.06;0.44] & 4.06 [1.12;11.81] \\
    2 & 2 & 0.61 [0.52;0.71] & 0.07 [0.04;0.11] & 1.12 [0.75;1.77] & 0.14 [0.07;0.45] & 5.27 [1.33;15.58] \\
    \bottomrule
  \end{tabular}
\end{sidewaystable}

\subsection{Robustness to the covariate dimension}\label{sm:sim-highdim}

\begin{figure}[!tb]
  \centering
  \includegraphics[width=0.9\textwidth]{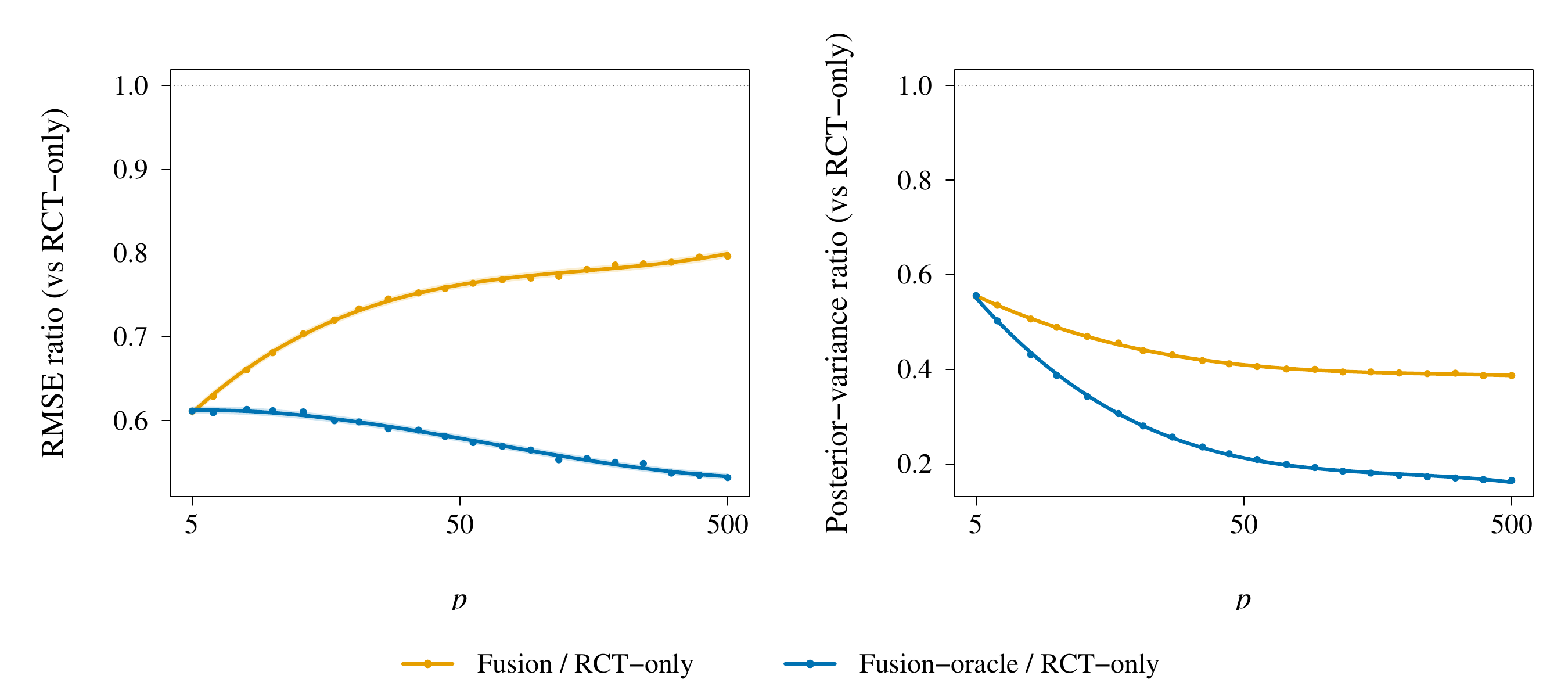}
  \caption{Efficiency of the Bayesian fusion forest relative to the trial-only baseline as the covariate dimension $p$ grows, at $\lambda_d = \lambda_u = 1$. Left: ratio of RMSE; right: ratio of posterior variance. A ratio below one favours the fusion. The fusion oracle is the Bayesian fusion forest given only the active covariates. Shaded bands are $95\%$ Monte Carlo intervals.}
  \label{fig:sim-hd-ratio}
\end{figure}

We report the efficiency of the fusion relative to the trial-only baseline as the covariate dimension $p$ grows.
Figure~\ref{fig:sim-hd-ratio} shows the ratio of the fusion to the trial-only RMSE and posterior variance, alongside the oracle that is handed only the active covariates.
Both ratios stay below one across the whole range, so the fusion improves on the trial-only analysis at every $p$.
The RMSE ratio drifts upward as $p$ grows, because the fusion must also learn the high-dimensional baseline, deviation, and confounding surfaces.
The oracle ratio instead falls, which confirms that this erosion is the cost of searching the noise covariates rather than a failure of borrowing.
The posterior-variance ratio decreases with $p$, so the fusion's precision advantage over the trial-only baseline in fact widens.

\subsection{The nonparametric error distribution}\label{sm:sim-error}

We assess the source-specific hierarchical Dirichlet process mixture (HDPM) that the Bayesian fusion forest places on the residual.
We isolate the residual model by setting $\lambda_u = \lambda_d = 0$, so the two sources share the same structural mean and differ only in their error distribution.
We compare three residual priors within the Bayesian fusion forest: a single Gaussian, a Dirichlet process mixture pooled across the two sources (one residual law for both), and the proposed source-specific HDPM (shared atoms, source-specific weights and scale).
We escalate the residual difficulty over five settings.
Setting~1 draws both errors from the standard normal, so the Gaussian prior is correctly specified.
Settings~2--5 give the real-world source a markedly larger scale than the trial (a standard-deviation ratio of about three), which the single-scale Gaussian and pooled priors cannot represent, and add increasing non-Gaussianity: a pure scale gap (2), a skewed real-world error (3), a bimodal real-world error (4), and multimodal errors in both sources with shared component locations but different weights (5).
To make the residual model bite, we censor heavily and coarsely: the trial is right-censored at circa $50\%$ and the real-world data are interval-censored at quartile visits, so many observations enter the likelihood only through the error distribution over wide intervals.
We report the CATE $\tau(x)$, our estimand throughout.

Table~\ref{tab:sim-err} reports the CATE metrics over the combined population.
At the correctly-specified null the three priors are indistinguishable, so the flexible residual models cost nothing when a Gaussian would do.
From Setting~2 onward the HDPM attains the lowest RMSE at every setting, together with the lowest, or tied-lowest, bias.
The Gaussian prior is the most biased throughout, and its bias grows as the error departs from normality, because under heavy coarse censoring a misspecified residual distribution biases the imputation of the censored events.
The pooled mixture removes much of this bias by modelling the shape flexibly, but, tied to a single residual law and scale for both sources, it cannot match the source-specific HDPM once the sources differ.
Coverage is near nominal, and slightly conservative, for all three priors, and posterior variance is comparable, so the accuracy gains of the HDPM do not come at a calibration cost.
The pattern is the same in the trial- and real-world-only populations.

\begin{table}[!htbp]
  \centering
  \caption{CATE metrics by residual prior across the five error settings, over
    the combined population and averaged over simulation replicates. ``Source
    HDPM'' is the proposed source-specific hierarchical Dirichlet process
    mixture; ``Shared DP'' pools one Dirichlet process mixture across sources.
    The best RMSE and bias per setting are in bold. Coverage is the $95\%$
    credible-interval coverage; ``Post.\ var.'' is the mean posterior variance
    of $\hat\tau(x)$.}
  \label{tab:sim-err}
  \small
  \begin{tabular}{@{}l rrrr@{}}
    \toprule
    Residual prior & RMSE & Bias & Coverage & Post.\ var. \\
    \midrule
    \multicolumn{5}{l}{\emph{Setting 1 --- Gaussian, identical (null)}}\\
    \quad Gaussian   & 0.745          & $-0.099$          & 0.956 & 0.557 \\
    \quad Shared DP  & 0.744          & $\mathbf{-0.064}$ & 0.957 & 0.561 \\
    \quad Source HDPM & $\mathbf{0.743}$ & $-0.066$        & 0.959 & 0.576 \\
    \addlinespace
    \multicolumn{5}{l}{\emph{Setting 2 --- Gaussian, per-source scale gap}}\\
    \quad Gaussian   & 0.751          & $-0.146$          & 0.964 & 0.627 \\
    \quad Shared DP  & 0.745          & $-0.111$          & 0.964 & 0.629 \\
    \quad Source HDPM & $\mathbf{0.739}$ & $\mathbf{-0.099}$ & 0.965 & 0.634 \\
    \addlinespace
    \multicolumn{5}{l}{\emph{Setting 3 --- Skewed RWD, scale gap}}\\
    \quad Gaussian   & 0.757          & $-0.179$          & 0.961 & 0.613 \\
    \quad Shared DP  & 0.745          & $\mathbf{-0.119}$ & 0.962 & 0.608 \\
    \quad Source HDPM & $\mathbf{0.740}$ & $-0.121$        & 0.963 & 0.613 \\
    \addlinespace
    \multicolumn{5}{l}{\emph{Setting 4 --- Bimodal RWD, scale gap}}\\
    \quad Gaussian   & 0.758          & $-0.138$          & 0.965 & 0.655 \\
    \quad Shared DP  & 0.754          & $-0.117$          & 0.966 & 0.666 \\
    \quad Source HDPM & $\mathbf{0.742}$ & $\mathbf{-0.096}$ & 0.967 & 0.661 \\
    \addlinespace
    \multicolumn{5}{l}{\emph{Setting 5 --- Multimodal, shared atoms, weight + scale gap}}\\
    \quad Gaussian   & 0.771          & $-0.150$          & 0.966 & 0.691 \\
    \quad Shared DP  & 0.768          & $-0.136$          & 0.967 & 0.706 \\
    \quad Source HDPM & $\mathbf{0.752}$ & $\mathbf{-0.107}$ & 0.968 & 0.694 \\
    \bottomrule
  \end{tabular}
\end{table}


\newpage
\section{Additional results for the data analysis}\label{sm:analysis-additional}

We collect here the material supporting the data analysis of the main text.
Section~\ref{sm:analysis-cohort} describes the analysis cohort and compares the Bayesian fusion forest with the trial-only fit.
Section~\ref{sm:analysis-competitors} reports the full comparison with the machine-learning alternatives.

\subsection{The analysis cohort and the trial-only comparison}\label{sm:analysis-cohort}

\begin{figure}[!tb]
  \centering
  \includegraphics[width=0.8\linewidth]{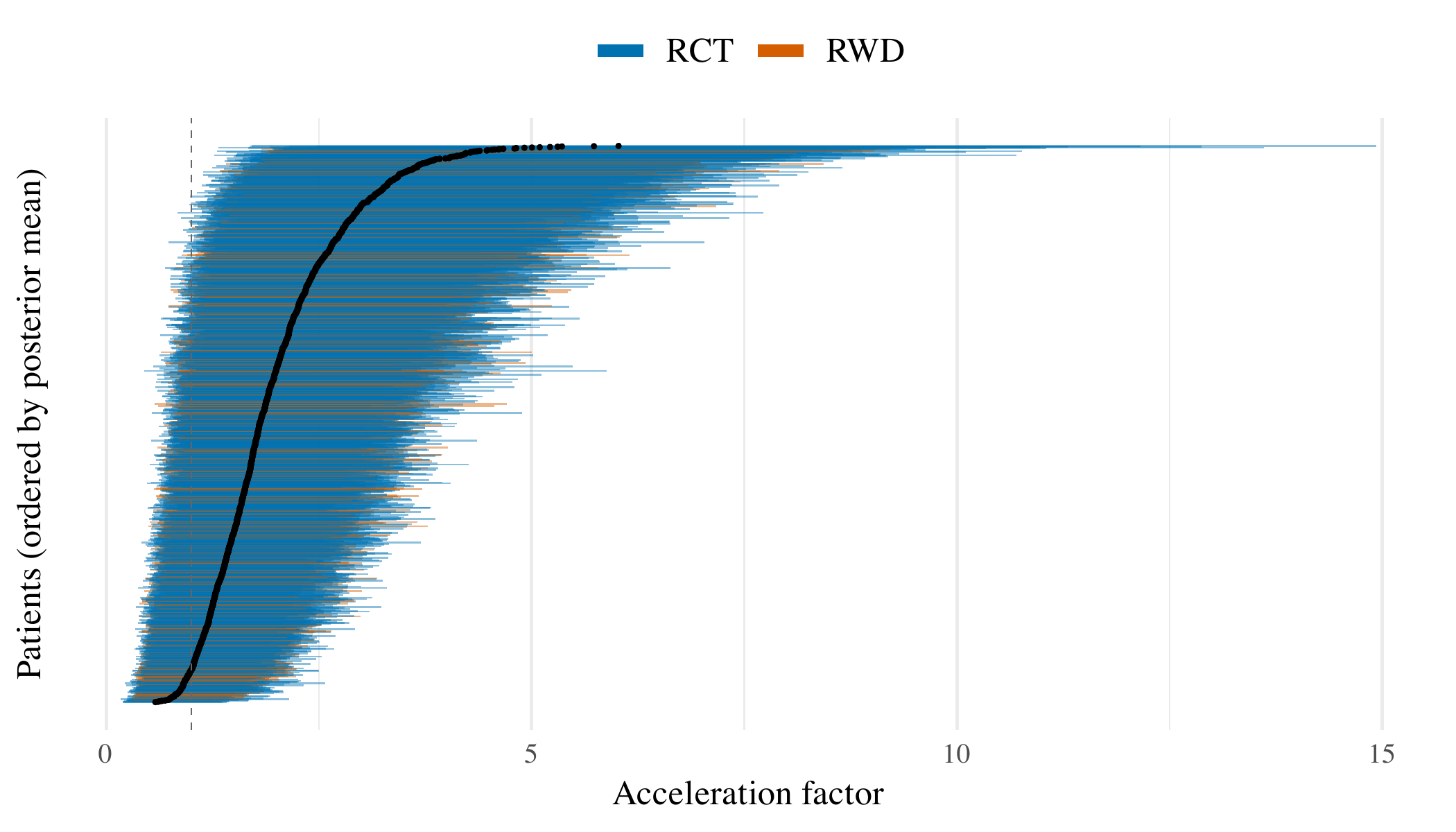}
  \caption{Subject-level acceleration factor for every patient in the
    trial-aligned cohort, from the trial-only analysis, ordered by
    posterior mean. Vertical bars are pointwise 95\% credible intervals;
    the dashed line at $1$ marks no effect. The intervals are wider than
    those of the Bayesian fusion forest in Figure~\ref{fig:cate} of the main text. }
  \label{fig:cate-rct}
\end{figure}

Table~\ref{tab:cohort} characterises the trial-aligned analysis cohort.
Figure~\ref{fig:cate-rct} gives the subject-level acceleration factor from the trial-only analysis, the comparator for the fusion estimates in the main text.
Table~\ref{tab:precision} compares the precision of the subject-level estimates between the trial-only and fusion analyses.

\begin{table}[h]
  \centering
  \caption{Characteristics of the trial-aligned analysis cohort of
    $2144$ patients, by data source. Continuous variables are reported
    as median [IQR], categorical variables as $n$ (\%).}
  \label{tab:cohort}
  \begin{tabular}{@{}lll@{}}
    \toprule
                                    & ACTG~175 (RCT)  & MACS (RWD)        \\
    \midrule
    Age, y                          & 34 [29;41]     & 39 [34;44]      \\
    CD4, cells/mm$^3$               & 337 [263;422]  & 263 [178;365]   \\
    CD8, cells/mm$^3$               & 918 [670;1246] & 849 [659;1153]  \\
    Calendar year                   & 1992            & 1991 [1991;1992]\\
    Prior ART, y                    & 0.4 [0;2.1]    & 1.0 [0;1]       \\
    Non-white, $n$ (\%)             & 404 (23)        & 78 (21)          \\
    \bottomrule
  \end{tabular}
\end{table}

\begin{table}[h]
  \centering
  \caption{Precision of the subject-level acceleration factor over the
    combined cohort, under the trial-only and fusion analyses. Both
    quantities are averaged across patients on the acceleration-factor
    scale.}
  \label{tab:precision}
  \begin{tabular}{@{}lrrr@{}}
    \toprule
                                & Trial-only & Fusion & Reduction \\
    \midrule
    Average 95\% credible-interval width & $3.16$  & $1.21$  & $62\%$ \\
    Average posterior variance  & $0.808$ & $0.098$ & $88\%$ \\
    \bottomrule
  \end{tabular}
\end{table}

\subsection{Comparison with the machine-learning alternatives}\label{sm:analysis-competitors}

We compare the results of the Bayesian fusion forest on this cohort with those of four flexible machine-learning predictors for survival outcomes: a deep accelerated failure time network \citep{norman2024deepaft}, gradient boosting under an accelerated failure time loss \citep{barnwal2022xgboost}, and Buckley--James boosting over extreme learning machines and over regression trees \citep{kong2023bjelm}.
We fit each as an S-learner and a T-learner \citep{kunzel2019metalearners}, on the trial alone and on a naive pool of both sources with a source indicator.
This gives sixteen configurations in total.
None of these methods have intrinsic uncertainty quantification, so we obtain intervals from $100$ stratified bootstrap resamples in every configuration.

The neural network architecture matches the simulation study.
We tune its learning rate, momentum, and weight decay by ten-fold cross-validation on the concordance index.
The boosted trees are tuned by random search with five-fold cross-validation, following the protocol of their authors; the two Buckley--James methods use their authors' default settings.
The network and both Buckley--James methods require right-censored data.
We therefore approximate the interval-censored MACS deaths by their right-censored last-observation encoding, the same encoding used by the trial-only forest.
This approximation discards the interval information.
The accelerated failure time loss of the boosted trees accepts interval labels, so that method is fitted on the actual bounds.

The cohort carries no ground truth, so nothing here reveals which method is closest to the true effect.
Interval width in particular is not a measure of quality but of uncertainty, and it carries that meaning only if the method producing it is calibrated.
For a calibrated method a narrower interval is genuinely informative, in that it permits a firmer conclusion about the same patient.
For a miscalibrated one it merely understates what is not known, and this cohort cannot tell the two cases apart.
In the simulation study, where the true conditional effect is known and coverage can therefore be measured, we showed that this distinction is not academic (Table~\ref{tab:sim-competitors}).
Every S-learner covered the truth between $0.00$ and $0.45$ of the time against a nominal $0.95$, while carrying intervals narrower than the Bayesian fusion forest's in every case.
Every T-learner reached between $0.46$ and $0.99$ coverage, but only at widths of $1.25$ to $5.97$, against $2.82$ for the Bayesian fusion forest, and at a higher error.
No configuration of any method achieved accurate estimation and calibrated intervals together, and the lowest root mean squared error of any of them, $1.01$, remained half again the Bayesian fusion forest's $0.68$.

\begin{figure}[!tb]
  \centering
  \setlength{\tabcolsep}{1pt}
  \renewcommand{\arraystretch}{0.6}
  \begin{tabular}{@{}c@{\hspace{2pt}}cccc@{}}
    & \small Trial, S-learner & \small Trial, T-learner
    & \small Pool, S-learner & \small Pool, T-learner \\
    \rotatebox{90}{\small\quad DNN} &
      \includegraphics[width=0.235\textwidth]{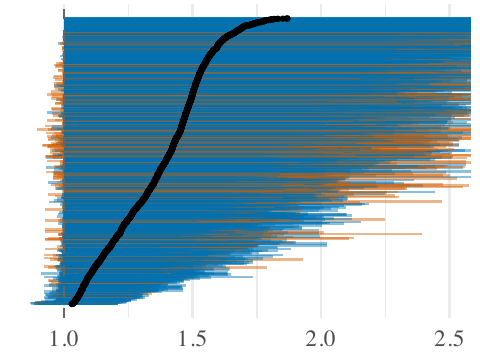} &
      \includegraphics[width=0.235\textwidth]{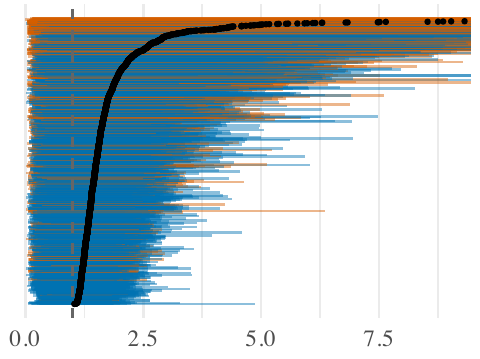} &
      \includegraphics[width=0.235\textwidth]{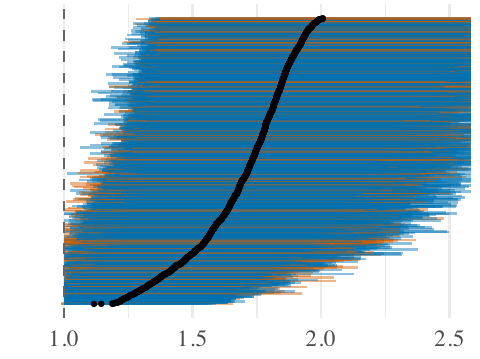} &
      \includegraphics[width=0.235\textwidth]{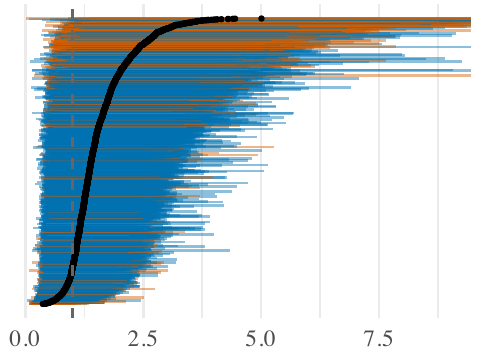} \\
    \rotatebox{90}{\small XGBoost} &
      \includegraphics[width=0.235\textwidth]{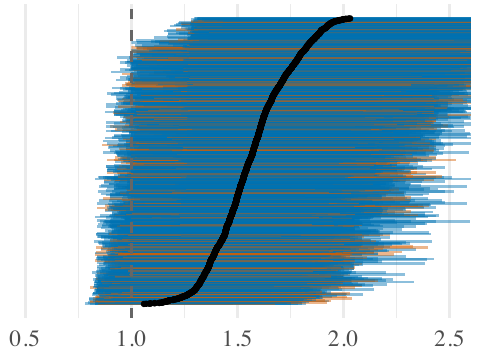} &
      \includegraphics[width=0.235\textwidth]{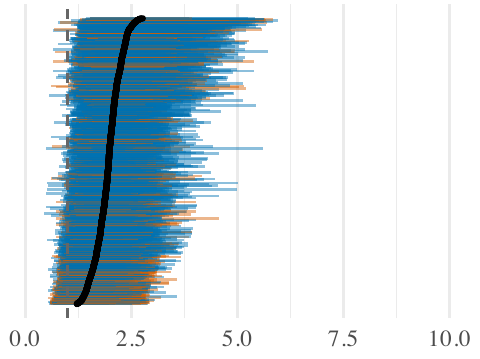} &
      \includegraphics[width=0.235\textwidth]{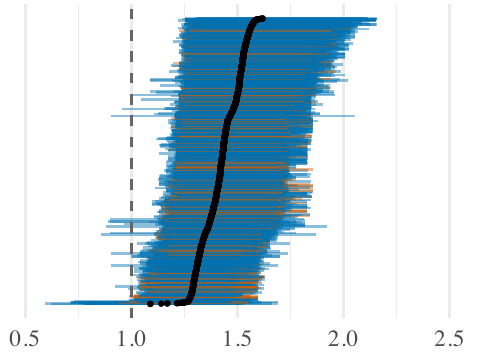} &
      \includegraphics[width=0.235\textwidth]{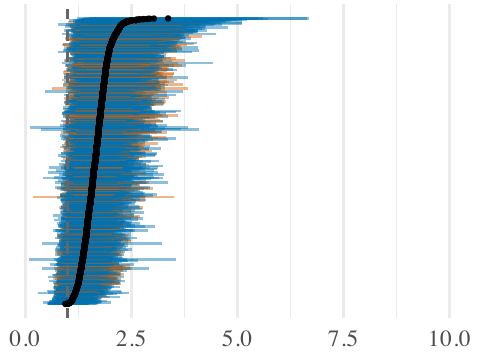} \\
    \rotatebox{90}{\small\, BJ-ELM} &
      \includegraphics[width=0.235\textwidth]{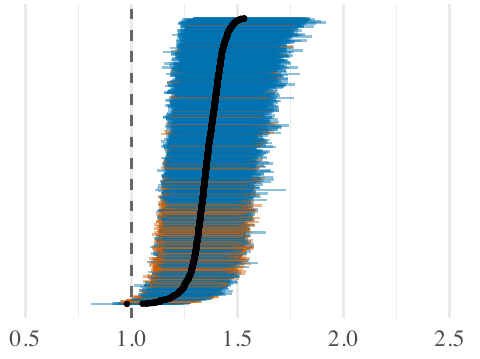} &
      \includegraphics[width=0.235\textwidth]{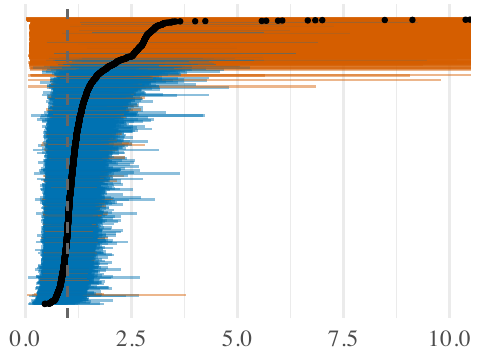} &
      \includegraphics[width=0.235\textwidth]{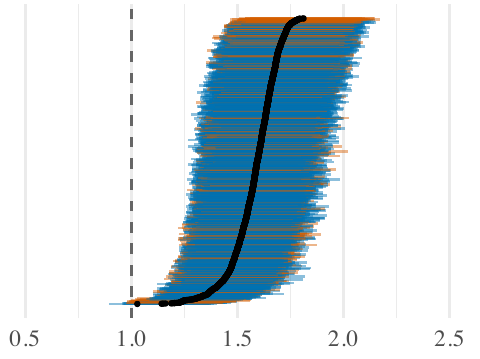} &
      \includegraphics[width=0.235\textwidth]{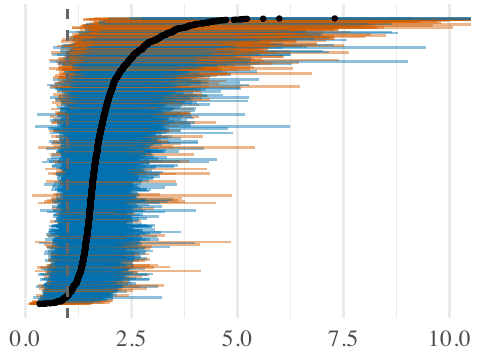} \\
    \rotatebox{90}{\small BJ-trees} &
      \includegraphics[width=0.235\textwidth]{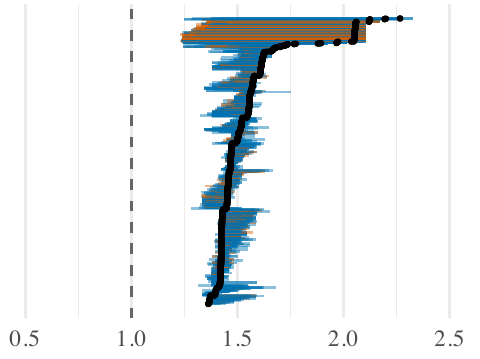} &
      \includegraphics[width=0.235\textwidth]{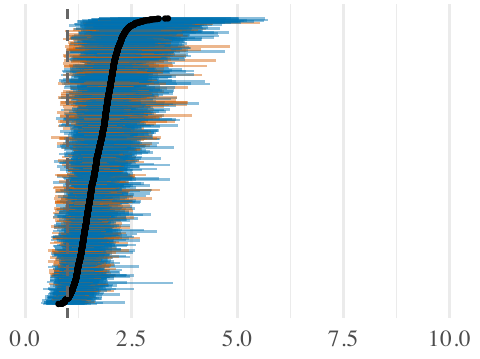} &
      \includegraphics[width=0.235\textwidth]{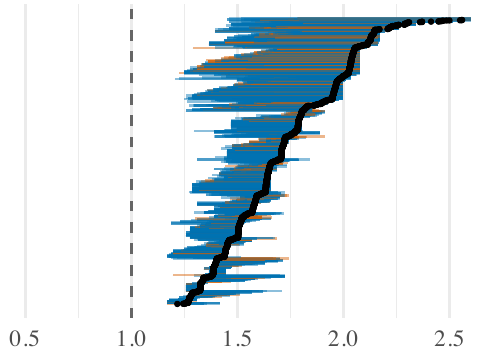} &
      \includegraphics[width=0.235\textwidth]{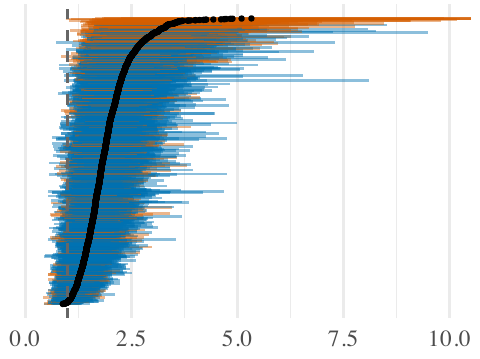} \\
  \end{tabular}
  \caption{Subject-level acceleration factor from every machine-learning
    alternative over the combined RCT + MACS cohort, by method (rows) and by
    training design and learner (columns). Patients are ordered by the
    bootstrap mean; bars are the $95\%$ bootstrap interval and the horizontal
    axis is the acceleration factor, with the dashed line at $1$ marking no
    effect. Colour indicates the data source, blue for the trial and vermillion
    for the cohort. Every S-learner column compresses the patients into a
    narrow band; every T-learner column scatters them from below one to above
    four. Compare the single calibrated fusion caterpillar of
    Figure~\ref{fig:cate} in the main text.}
  \label{fig:cate-competitors}
\end{figure}

Table~\ref{tab:sm-analysis-competitors} and Figure~\ref{fig:cate-competitors} show that the cohort reproduces the same two shapes, method for method.
Every S-learner compresses the patients into a narrow band, with an interquartile range of the per-patient estimate between $0.11$ and $0.39$ on the acceleration-factor scale.
Every T-learner scatters them, with an interquartile range of $0.37$ to $0.69$ and individual estimates reaching from $0.02$ to $7.65$ at the extremes.
Because the simulation established that these narrow S-learner intervals undercover badly, the narrow intervals here are to be read as overconfidence rather than precision.
The wide T-learner intervals are the mirror image: the simulation showed they buy their coverage with width, and here they reach more than four times the Bayesian fusion forest's.
The consequence for a clinical reading is severe.
The share of patients declared at least $95\%$ certain to benefit ranges from $0.0\%$ to $100\%$ across the sixteen configurations.
Within a single method it can move almost the whole way: the deep network declares no patient certain to benefit when trained on the trial and $96.1\%$ when trained on the pool, and Buckley--James boosting over extreme learning machines declares $99.1\%$ as an S-learner and $3.3\%$ as a T-learner on the very same data.
These are not differences of degree but opposite clinical conclusions, and neither the data nor any principled rule selects between them.

The Bayesian fusion forest is not exposed to any of this.
It is not a meta-learner, so there is no S-versus-T choice to make and no second configuration that could have reversed the answer.
It targets the causal contrast directly rather than differencing two predictions, and it combines the two sources through the confounding function rather than by pooling them and hoping.
Its intervals are posterior intervals, so they propagate the uncertainty of the fit instead of resampling around it, and the simulation confirmed they are calibrated: coverage $0.96$ at the lowest error of any method considered.
On this cohort it returns one answer, with a $95\%$ interval of average width $1.21$ and $96.4\%$ of patients at least $95\%$ certain to benefit.
That single number is not more comfortable than the alternatives; it is the only one that comes with evidence that it can be trusted.

\begin{table}[!tb]
  \centering
  \small
  \begin{tabular}{@{}lll rr@{}}
    \toprule
    Method & Source & Learner & Mean $95\%$ width & $\geq 95\%$ certain \\
           &        &         & (AF scale)        & of benefit (\%)     \\
    \midrule
    Bayesian fusion forest   & Both  & --- & 1.21 & 96.4 \\
    Trial-only causal forest & Trial & --- & 3.15 & 36.8 \\
    \midrule
    Deep neural network      & Trial & S & 1.06 & 0.0 \\
                             &       & T & 5.09 & 0.1 \\
                             & Pool  & S & 1.16 & 96.1 \\
                             &       & T & 3.32 & 6.6 \\
    \addlinespace
    XGBoost                  & Trial & S & 1.20 & 60.8 \\
                             &       & T & 2.14 & 76.5 \\
                             & Pool  & S & 0.57 & 98.8 \\
                             &       & T & 1.54 & 69.7 \\
    \addlinespace
    BJ-ELM                   & Trial & S & 0.41 & 99.1 \\
                             &       & T & 4.07 & 3.3 \\
                             & Pool  & S & 0.53 & 99.9 \\
                             &       & T & 2.39 & 48.7 \\
    \addlinespace
    BJ-trees                 & Trial & S & 0.16 & 100.0 \\
                             &       & T & 1.31 & 70.8 \\
                             & Pool  & S & 0.32 & 100.0 \\
                             &       & T & 1.85 & 68.1 \\
    \bottomrule
  \end{tabular}
  \caption{The machine-learning alternatives over the combined RCT + MACS
    cohort ($n = 2144$). ``Mean $95\%$ width'' is the average width of the
    per-patient interval on the acceleration-factor scale. The last column is
    the percentage of patients at least $95\%$ certain to benefit; for the
    alternatives this is a bootstrap proportion rather than a posterior
    probability. The network and the two Buckley--James methods receive the
    right-censored approximation of the interval-censored MACS deaths, while
    XGBoost is fitted on the interval bounds. The Bayesian fusion forest and the
    trial-only forest are not meta-learners and have a single configuration
    each. Neither column measures accuracy: the cohort has no ground truth, and
    the calibration of these intervals is established in the simulation study
    of Table~\ref{tab:sim-competitors}.}
  \label{tab:sm-analysis-competitors}
\end{table}

\end{document}